\documentclass[acmsmall]{acmart}
\usepackage{tcolorbox}
\usepackage{fbox}
\usepackage{enumitem}
\usepackage{graphicx}
\usepackage{xcolor}

\usepackage{algorithm}
\usepackage{algpseudocode}

\usepackage{soul,xcolor}
\setstcolor{red}

\algrenewcommand\algorithmicensure{\textbf{Return:}}
\definecolor{ballblue}{rgb}{0.13, 0.67, 0.8}

\AtBeginDocument{%
  \providecommand\BibTeX{{%
    \normalfont B\kern-0.5em{\scshape i\kern-0.25em b}\kern-0.8em\TeX}}}

\setcopyright{rightsretained}
\acmDOI{10.1145/3660805}
\acmYear{2024}
\copyrightyear{2024}
\acmSubmissionID{fse24main-p752-p}
\acmJournal{PACMSE}
\acmVolume{1}
\acmNumber{FSE}
\acmArticle{98}
\acmMonth{7}
\received{2023-09-28}
\received[accepted]{2024-04-16}

\begin{document}

\title{BARO: Robust Root Cause Analysis for Microservices via Multivariate Bayesian Online Change Point Detection}

\author{Luan Pham}
\orcid{0000-0001-7243-3225}
\affiliation{%
  \institution{RMIT University}
  % \city{Melbourne}
  \country{Australia}
}
\email{luan.pham@rmit.edu.au}

\author{Huong Ha}
\orcid{0000-0003-2463-7770}
\affiliation{%
  \institution{RMIT University}
  % \city{Melbourne}
  \country{Australia}
}
\email{huong.ha@rmit.edu.au}

\author{Hongyu Zhang}
\orcid{0000-0002-3063-9425}
\affiliation{%
  \institution{Chongqing University}
  % \city{Chongqing}
  \country{China}
}
\email{hyzhang@cqu.edu.cn}

\begin{abstract}

Detecting failures and identifying their root causes promptly and accurately is crucial for ensuring the availability of microservice systems. A typical failure troubleshooting pipeline for microservices consists of two phases: anomaly detection and root cause analysis. While various existing works on root cause analysis require accurate anomaly detection, there is no guarantee of accurate estimation with anomaly detection techniques. Inaccurate anomaly detection results can significantly affect the root cause localization results. To address this challenge, we propose \textit{BARO}, an end-to-end approach that integrates anomaly detection and root cause analysis for effectively troubleshooting failures in microservice systems. BARO leverages the Multivariate Bayesian Online Change Point Detection technique to model the dependency within multivariate time-series metrics data, enabling it to detect anomalies more accurately. BARO also incorporates a novel nonparametric statistical hypothesis testing technique for robustly identifying root causes, which is less sensitive to the accuracy of anomaly detection compared to existing works. Our comprehensive experiments conducted on three popular benchmark microservice systems demonstrate that BARO consistently outperforms state-of-the-art approaches in both anomaly detection and root cause analysis.

\end{abstract}
% \keywords{Root Cause Analysis, Microservices, AIOps}




\maketitle

\section{Introduction}
\label{sec:introduction}
In recent years, microservice systems have gained significant popularity in the development of cloud-based applications, owing to their numerous advantages such as resource flexibility, a loosely coupled architecture, and lightweight deployment. However, failures are inevitable in microservice systems due to their inherent complexity. A failure in one service can propagate across the system, affecting many other services and resulting in the degradation of the system availability. This, in turn, leads to poor user experience and incurs huge economic losses. For instance, it has been reported that a one-hour downtime on Amazon.com could potentially cost up to 100 million USD~\cite{Chen2019incienttriage, Chen2020incidentmanagment}. Therefore, system operators must closely monitor the systems, checking key run-time information to promptly detect failures as soon as they occur, and then proceed to identify the failures' root causes and troubleshoot them. However, in practice, the complexity of microservice systems and the large volume of monitoring data make these tasks especially challenging.

Metric-based anomaly detection and root cause analysis (RCA) for microservice systems have been extensively studied in recent years \cite{Soldani2022rcasurvey, Azam2022rcd, Li2022Circa, Xin2023CausalRCA, chen2022adaptive, siffer2017anomaly, wu2020microrca, Jinjin2018Microscope, yu2021microrank, liu2021microhecl,lee2023eadro}. Given a set of metrics data, anomaly detection techniques aim to detect whether there exist anomalies and consequently, failures within the microservice system~\cite{chen2022adaptive, lee2023eadro}. If a failure is detected, the RCA module is then triggered to locate the root cause of the failure \cite{lee2023eadro, Azam2022rcd, Li2022Circa}. The RCA module aims to address two fundamental questions: (1) which services are the root causes, and (2) what specific issues are causing that failure (e.g. high CPU utilization, memory leak, or network congestion). Some RCA approaches use hypothesis testing or a statistical analysis method to analyze the time series metrics data to identify the candidate root causes for the detected anomalies \cite{Shan2019Ediagnosis, Li2022Circa}. Some RCA methods construct topology graphs using the information provided by the monitoring systems, such as the microservice status, the interaction between the services, and the interaction traces, to facilitate root cause analysis \cite{wu2020microrca, liu2021microhecl}. Multiple recent RCA methods \cite{Xin2023CausalRCA, Azam2022rcd, Jinjin2018Microscope, Meng2020Microcause} use different causal discovery methods \cite{Runge2019PCMCI, Jaber2020PsiFCI, Spirtes1995fci} to derive the causal relationships among the services and metrics from the multivariate time series metrics data and employ graph centrality algorithms like random walk \cite{Jinjin2018Microscope}, PageRank \cite{Xin2023CausalRCA,wu2021microdiag} or Depth-First Search (DFS) \cite{Chen2014Causeinfer} to infer the root causes.

Despite being closely related, existing RCA works typically either treat the anomaly detection tasks independently or rely on overly simplistic anomaly detection techniques~\cite{lee2023eadro, Soldani2022rcasurvey, Li2022Circa, Azam2022rcd}. For example, MicroRCA \cite{wu2020microrca} and MicroDiag \cite{wu2021microdiag} employ the simple BIRCH clustering technique~\cite{zhang1996birch} for detecting anomalies. MicroScope \cite{Jinjin2018Microscope} and Ms-Rank \cite{Ma2019Msrank} use a basic three-sigma rule of thumb as their anomaly detection method, known as N-Sigma. 
There are various commercial monitoring platforms, notable platforms including DataDog \cite{datadog} and Dynatrace \cite{dynatrace}, which rely on simplistic %threshold-based 
anomaly detection techniques \cite{dynatrace1, datadog2} for univariate time series data. 
Most RCA research works~\cite{Jinjin2018Microscope, Wang2018cloudranger, Ma2019Msrank, wu2020microrca, Ma2020Automap, wu2021microdiag, Li2022Circa, Azam2022rcd, Xin2023CausalRCA} focus solely on identifying the root cause of the failure whilst assuming the existence of an anomaly detection module that can accurately detect failures and trigger the RCA module when failures are detected. Some of these works, such as CIRCA \cite{Li2022Circa} and RCD \cite{Azam2022rcd}, specifically require certain information from the anomaly detection module, in particular, the failure occurrence time. However, they assume this information is already known accurately. Thus, it remains unclear whether, when combined with existing anomaly detection methods that may provide imprecise information, these approaches are still effective in localizing the root cause.

In this work, we introduce BARO, an end-to-end approach for anomaly detection and root cause analysis for microservice systems based on metrics data, which are multivariate time series data. BARO includes a Multivariate Bayesian Online Change Point Detection module for detecting anomalies. It also includes a novel RobustScorer module, which is a nonparametric statistical hypothesis testing technique and less sensitive to the accuracy of the anomaly detection, for robustly identifying the root causes. BARO offers several advantages. Firstly, it follows an unsupervised learning approach, eliminating the need for labelled data and enabling direct application without the requirement of such labelled data. Secondly, it does not rely on operational knowledge (e.g., service call graphs) or causal graphs, making it suitable for large-scale evolving systems where acquiring such operational knowledge or causal graphs for numerous services is difficult \cite{Azam2022rcd, Li2022Circa, Wang2018cloudranger}. Finally, BARO is nonparametric, scale-equivalent, and rotation-invariant, making it applicable to a wide range of systems. We comprehensively evaluate BARO against various state-of-the-art approaches on three popular benchmark microservice systems. Our experimental results demonstrate that BARO consistently surpasses the state-of-the-art methods. Additionally, we analyze the sensitivity of the RCA methods against their parameters to show the robustness of our method.

In summary, our major contributions are as follows: 
\begin{itemize}
    \item We propose a new end-to-end approach for anomaly detection and root cause analysis in microservice systems based on multivariate time series metrics data. In particular, we propose to use the Multivariate Bayesian Online Change Point Detection technique to detect anomalies, and a novel nonparametric statistical hypothesis testing technique for %robustly 
    accurately identifying root causes of microservice systems' failures.
    \item We conduct extensive experiments on three popular benchmark microservice systems. Our experimental results demonstrate that BARO consistently outperforms state-of-the-art approaches in both anomaly detection and root cause analysis.
    \item We perform a comprehensive sensitivity analysis, evaluating the performance of all the RCA methods w.r.t. their parameters. Our experimental results show that BARO is significantly more robust against important parameters% compared to baseline methods.
    , e.g., the anomaly detection time, compared to baseline methods.
\end{itemize}

\section{Problem Statement and Background}
\label{sec:background}
\subsection{Problem Statement}

\subsubsection{Key Terminology} \label{sec:key-terms}

\textit{Failures} represent the actual inability of a service to execute its functions~\cite{Soldani2022rcasurvey}. \textit{Faults} correspond to the root causes of such failures (e.g., CPU hog, memory leak, or network disconnection) \cite{Soldani2022rcasurvey, avizienis2004basic}. \textit{Anomalies} are defined as observable symptoms of failures~\cite{Li2022Circa, Soldani2022rcasurvey}. \textit{Root cause analysis (RCA)} is the process of determining why a failure has occurred~\cite{lee2023eadro}, i.e., finding the root cause of the failure. RCA involves a thorough examination of various monitoring data, i.e., including metrics data. \textit{Metrics} are recorded by the monitoring system and contain various critical information within the microservice systems, such as workload, resource consumption, and response time \cite{Xin2023CausalRCA}. These metrics are typically represented as multivariate time series, with each time series corresponding to the data collected with a specific metric. In the context of metric-based RCA, \textit{root cause metrics} are the metrics that are indicators of the root cause \cite{Li2022Circa, chen2022adaptive, Azam2022rcd}. The system operators can use these suggested root cause metrics to identify the true underlying root cause of the failures. The use of these terms aligns with existing RCA works \cite{Chen2014Causeinfer, chen2016causeinfer, chen2022adaptive, Azam2022rcd, Li2022Circa, liu2023pyrca, Meng2020Microcause, Xin2023CausalRCA}.

\subsubsection{Problem Formulation} \label{sec:problem-formulation}
Let us consider a microservice system $\mathcal{S}$ consisting of $n$ services $\{ s^i \}_{i=1}^n$. At each time step $t$, the monitoring system collects $m$ metrics $\mathcal{M}^{i,j=1:m}_t = \{x^{(i, j)}_t\}_{j=1}^{m} (m \geq 1)$ from each service $s^i$. Given a $T$-length observation window with time-series metrics data $\mathcal{M}^{i=1:n,j=1:m}_{t_0:t_0+T}$, our goal is to develop a framework that solves two problems. The first problem is to \textbf{\textit{predict the existence of anomalies}} (failures), represented as a binary indicator $y$, which takes the value of $0$ when there is no anomaly and $1$ when there is an anomaly within the metrics dataset. The second problem is that, when $y$ returns $1$, an RCA module is triggered to \textit{\textbf{pinpoint the root causes of the failure}} using this dataset $\mathcal{M}^{i=1:n,j=1:m}_{t_0:t_0+T}$, i.e., the specific root cause services and the corresponding root cause metrics.

\subsection{Multivariate Time Series Data in Microservice Systems}

Metric-based anomaly detection and RCA are typically based on runtime information collected on the services within the microservice system. Such information includes metrics monitored on the microservices, such as workload, resource consumption, and response time. These metrics are typically represented as multivariate time series, with each time series corresponding to the data collected with a specific metric \cite{Soldani2022rcasurvey}. Microservices also generate logs to provide more detailed and meaningful information about their state. Some previous studies~\cite{wang2021detecting, aggarwal2020localization, aggarwal2021causal} parse raw logs to extract log static structures (i.e., log templates \cite{le2023log}), and count the occurrences of these templates, which are subsequently transformed into time series data. These logs can also be a source of multivariate time series data, which can be used for analyze root cause. In this work, we, however, only focus on metrics as multivariate time series data, as with~\cite{Li2022Circa, Azam2022rcd, wu2020microrca, wu2021microdiag, thalheim2017sieve, Meng2020Microcause, Wang2018cloudranger, liu2021microhecl}.

\vspace{-0.1cm}
\subsection{Anomaly Detection} \label{sec:background-ad}

In microservice systems, once a failure occurs in a service, it is typically reflected in the metrics data of that particular service, resulting in an anomaly or a change in its data distribution. Furthermore, a failure in one service can propagate across the systems and impact other services, subsequently leading to changes to the metrics data of those services as well. To detect a failure in a microservice system, the goal is to detect any anomalies or changes in the given metrics dataset \cite{Soldani2022rcasurvey}.

A wide range of anomaly detection methods exist for time series data \cite{BlazquezGarc2021ADreview}. In this paper, since we target the problem of identifying root causes for microservice systems, we only focus on root cause localization-oriented anomaly detectors employed or discussed in existing metric-based RCA studies \cite{lee2023eadro}. N-Sigma \cite{Jinjin2018Microscope} is used and discussed in MicroRank \cite{yu2021microrank}, CIRCA \cite{Li2022Circa}, Eadro \cite{lee2023eadro}, and MicroScope \cite{Jinjin2018Microscope}. SPOT \cite{siffer2017anomaly} is mentioned and evaluated in CIRCA \cite{Li2022Circa} and Eadro \cite{lee2023eadro}.  BIRCH \cite{zhang1996birch} is employed in MicroRCA \cite{wu2020microrca}  and MicroDiag \cite{wu2021microdiag}. Finally, Univariate Offline Bayesian Change Point Detection \cite{adams2007bayesian} is used in CauseInfer \cite{Chen2014Causeinfer, chen2016causeinfer}. We describe these methods in the below.

\paragraph{N-Sigma}
N-Sigma (i.e., the three-sigma rule of thumb) represents one of the simplest anomaly detection methods. It operates based on the assumption that the data points falling within three standard deviations of the mean of the data distribution are considered normal. Consequently, any data point $x$ that falls outside this range, i.e., $\mu - 3\sigma < x < \mu + 3\sigma$, is deemed abnormal. For instance, MicroScope \cite{Jinjin2018Microscope} computes the mean and standard deviation of the metrics data distribution using the most recent 10 minutes of data and then applies this method to detect anomalies.

\vspace{-0.2cm}
\paragraph{SPOT}
SPOT and dSPOT \cite{siffer2017anomaly} are founded based on the principles of Extreme Value Theory \cite{Coles2001EVT} to detect anomalies and have been employed in various RCA research works \cite{pan2021dycauserca, Li2022Circa, lee2023eadro, Meng2020Microcause}. SPOT is designed to handle data with stationary distribution, while dSPOT is developed for streaming data susceptible to concept drift. It is worth noting that different studies have used different variants of SPOT; For instance, MicroCause \cite{Meng2020Microcause} and DycauseRCA \cite{pan2021dycauserca} made use of dSPOT, whereas CIRCA~\cite{Li2022Circa} employed biSPOT. In this work, we will employ dSPOT, as in \cite{Meng2020Microcause, pan2021dycauserca}. Therefore, in the sequel, when referring to SPOT, we are specifically referring to dSPOT. 

\vspace{-0.2cm}
\paragraph{BIRCH}
BIRCH (Balanced Iterative Reducing and Clustering using Hierarchies) \cite{zhang1996birch} is a well-regarded unsupervised clustering algorithm known for its efficiency in real-time data analysis and anomaly detection, particularly in large-scale datasets and time-series data. Its main idea is to generate a brief and informative summary about the original dataset, and then perform clustering on this summary. BIRCH considers a data point to be an anomaly when it is in a different cluster with other consecutive data points. MicroRCA \cite{wu2020microrca} and MicroDiag \cite{wu2021microdiag} have employed BIRCH for anomaly detection as a preliminary step before conducting root cause analysis.

\vspace{-0.2cm}
\paragraph{Univariate Offline Bayesian Change Point Detection}
Bayesian change point detection \cite{adams2007bayesian} is a statistical method for identifying change points in time series data, i.e., timesteps where the data distribution experiences significant shifts. It relies on the principle of causal predictive filtering, aiming to generate an accurate distribution of future unseen data points based solely past observations. Through Bayesian statistics, it incorporates the prior information regarding the characteristics of the change points into the modelling process, making it to be both effective and efficient. In previous RCA research works, such as \cite{Chen2014Causeinfer} and \cite{chen2016causeinfer}, univariate offline Bayesian change point detection was used to detect change points (anomalies) within the time series metrics data. 

Besides the methods mentioned, commercial platforms like Datadog \cite{datadog} and Dynatrace \cite{dynatrace} also provide anomaly detection techniques \cite{datadog2, dynatrace1} for univariate time series. These approaches either rely on users or historical data to obtain thresholds to detect anomalies. They are similar to the concept of N-Sigma, which uses expected values along with pre-defined tolerance thresholds.

% Besides the methods mentioned, commercial platforms like Datadog \cite{datadog} and Dynatrace \cite{dynatrace} also provide threshold-based anomaly detection techniques \cite{datadog2, dynatrace1} for univariate time series. These approaches rely on pre-defined thresholds to detect anomalies in incoming data, similar to the concept of N-Sigma, which uses expected values along with pre-defined tolerance thresholds.

\subsection{Root Cause Analysis}

Existing metric-based RCA algorithms can be classified into three main categories: statistical analysis, topology graph-based methods, and causal graph-based methods \cite{Soldani2022rcasurvey}.

\vspace{-0.2cm}
\paragraph{Statistical Analysis.} These approaches pinpoint failures' root causes by identifying metrics that undergo significant changes during the anomalous period. $\epsilon$-Diagnosis \cite{Shan2019Ediagnosis} uses the two-sample test algorithm and $\epsilon$-statistics to measure the similarity among the metrics and rank the root cause based on the similarity scores. $\epsilon$-Diagnosis is evaluated against three statistical analysis methods: Pearson distance, KNN \cite{friedman1983graph, luo2014correlating} and MST \cite{friedman1979multivariate}. KNN \cite{friedman1983graph, luo2014correlating} uses nearest neighbours to model the distance between two time series, while MST \cite{friedman1979multivariate} uses a minimum spanning tree to represent the distance. In \cite{wang2020root}, a neural network is used to learn the normal behaviour and measure the similarity of monitoring data when the failure happens. They use mutual information to rank the root causes. N-Sigma \cite{Jinjin2018Microscope,Li2022Circa} is another statistical analysis technique that assesses the distance using z-score.

\vspace{-0.2cm}
\paragraph{Topology Graph-based Analysis.} These approaches reconstruct a topology graph representing the microservice system using information from monitoring systems and operational knowledge. MicroRCA \cite{wu2020microrca} constructs a topology graph from monitoring data, extracts anomalous subgraphs, and uses the random walk algorithm to infer root causes. Similarly, \cite{wu2020performance} constructs a topology graph and anomalous subgraphs, followed by a neural network-based method to infer the root cause. Sieve \cite{thalheim2017sieve} uses a clustering technique to reduce the number of metrics on the constructed topology graph, then uses Granger Causality tests \cite{Granger1980causal} to determine the possible root causes. Likewise, Brandon \cite{brandon2020graph} augments the provided topology graph with monitored metrics, conducts root cause searches through extracted subgraphs, and ranks root causes based on similarity scores. DLA \cite{samir2019dla} transforms the provided topology graph and metric data into a hierarchical hidden Markov model and identifies root causes by computing the path with the highest anomalous probability. Meanwhile, CIRCA \cite{Li2022Circa} performs hypothesis testing on the structural graph to find the root causes. Commercial platforms such as Datadog \cite{datadog} and Dynatrace \cite{dynatrace} construct a topology graph from distributed traces and perform Depth First Search (DFS) to find the root cause services \cite{datadog1, dynatrace3}.

\vspace{-0.2cm}
\paragraph{Causal Graph-based Analysis.} 

Many recent RCA techniques adopt the causal graph-based approach \cite{Chen2014Causeinfer, chen2016causeinfer, Jinjin2018Microscope, Wang2018cloudranger, Chen2019airalert, Ma2019Msrank, Meng2020Microcause, Ma2020Automap, wu2021microdiag, Azam2022rcd, Xin2023CausalRCA}. The main idea is to construct a causal graph where vertices represent services or metrics of the microservices, and edges represent the cause-effect relationships between the services/metrics. These graphs are constructed using different causal discovery methods such as PC, FCI, LiNGAM, and GES \cite{Spirtes1993Causal, Spirtes1995fci, Chickering2002Ges, Shimizu2006Lingam, Jaber2020PsiFCI}. Assuming the root cause metric would affect other services' metrics, graph centrality algorithms like Breath First Search (BFS), random walk, or PageRank are used to rank the root causes. In addition, correlation analysis can be conducted to measure the correlation among metrics \cite{wu2021microdiag}, and the graph traversal process can consider these scores to identify the root cause. Recently, RCD \cite{Azam2022rcd} employs a divide-and-conquer strategy to split the input metrics into smaller chunks and constructs a causal graph for each chunk. It then employs $\Psi$-PC \cite{Jaber2020PsiFCI} to identify the root cause for each chunk, which is later combined to yield the final rootcause. CausalRCA \cite{Xin2023CausalRCA} introduces the use of a gradient-based causal discovery method, namely DAG-GNN, to uncover causal relationships among metrics.

\section{BARO: Proposed Root Cause Analysis Approach}
\label{sec:method}

In this section, we first introduce the basic assumptions underlying our approach (Section \ref{sec:assumption-causal}), then we present our proposed end-to-end approach for anomaly detection and root cause analysis, namely BARO (Sections \ref{sec:approach-ovw}, \ref{sec:multi-bocpd}, \ref{sec:robustscorer}). \textbf{BARO} incorporates a Multivariate \textbf{BA}yesian Online Change Point Detection technique to model the dependency and correlation structure of multivariate time series metrics data, enabling it to effectively detect anomalies and estimate the occurrence time of failures
(Section \ref{sec:multi-bocpd}). Then it uses a nonparametric hypothesis testing method, referred to as \textbf{RO}bustScorer, to reliably identify and rank the potential root causes, which is less sensitive to the accuracy of the anomaly detection time (Section \ref{sec:robustscorer}).

\subsection{Basic Assumptions} \label{sec:assumption-causal}

\subsubsection{Anomaly Metrics} \label{sec:anomaly-metric-selection}

As commonly recognized in previous works \cite{Li2022Circa, yu2023cmdiagnostor, liu2021microhecl}, there are generally four types of metrics: \textit{Traffic} (e.g., request count per minute), \textit{Saturation} (e.g., CPU utilization, database records), \textit{Latency} (e.g., average response time per minute), and \textit{Errors} (e.g., the rate of failed requests). These four types are named after the four golden signals in site reliability engineering \cite{googlesre}. Following the practice in metric-based RCA studies \cite{yu2023cmdiagnostor, Xin2023CausalRCA, Jinjin2018Microscope, Li2022Circa, Azam2022rcd}, we assume that \textit{anomalies should be visible in the metrics, subsequently affecting \textit{Latency} and/or \textit{Errors}.} With these assumptions, in varying conditions where there is a surge in \textit{Traffic} or \textit{Saturation} (e.g., in holiday periods) but without abnormal increases in \textit{Latency} and \textit{Errors}, our proposed method considers these situations as normal. Conversely, if a surge in \textit{Traffic} or \textit{Saturation} metrics causes abnormal increases in \textit{Latency} or \textit{Errors}, our method considers this an anomaly. Finally, for the RCA task, our method considers all the metrics to pursue the fine-grained output root cause ranking.

\subsubsection{Failure Propagation Chain} \label{sec:assumption-failure-propagation}

In practice, when a service failure occurs, it generally results in an anomaly in the data associated with the metric corresponding to that failure~\cite{Li2022Circa, Azam2022rcd}. For example, network congestion in a service typically leads to an increase in its response time. Furthermore, since a failure can propagate across the services within the system, this initial anomaly will then trigger additional anomalies in the metrics data of other services at later time~\cite{Azam2022rcd, Xin2023CausalRCA, Jinjin2018Microscope, Li2022Circa, liu2021microhecl}. Thus, when an anomaly is detected and the RCA module is activated, the anomalous period of runtime metrics data generally consists of multiple anomalies. In this work, based on this failure propagation chain, we assume \textit{the first anomaly corresponds to the time when the failure first occurs}. Thus, in our proposed method, we use the first detected anomaly to approximate the failure occurrence time ($\hat{t}_A$ in Alg. \ref{alg:baro}) to separate the normal and abnormal metrics data. It is important to note that this assumption does not imply the first detected anomaly to be the root cause, and, it has been implicitly used in previous RCA works \cite{Shan2019Ediagnosis, Li2022Circa, Azam2022rcd} for the same purpose as ours (i.e. to separate the abnormal and normal data). Our experimental results, along with those of previous studies, affirm the validity of this assumption.

\begin{figure}[t]
\includegraphics[width=\textwidth]{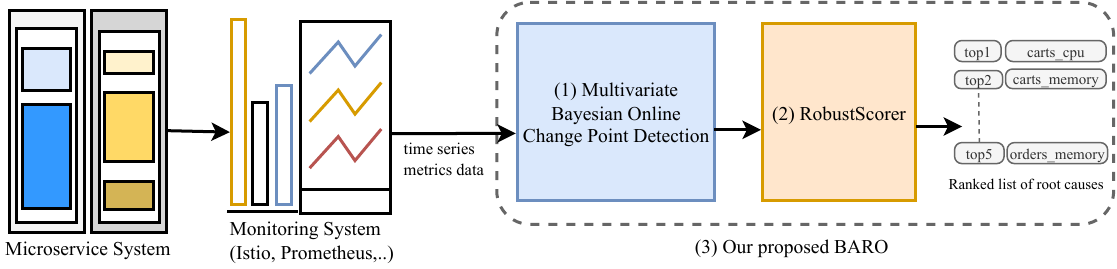}
\vspace{-0.5cm}
\caption{The overview: The monitoring system monitors the microservice system and collects the time series data. Our BARO consists of two components: Multivariate BOCPD and RobustScorer. The Multivariate BOCPD acts as an anomaly detection module to continuously check whether there is an anomaly. If there exists an anomaly, it triggers RobustScorer to score and rank the root cause services and metrics correspondingly.} \label{fig:intro2}
\end{figure}

\subsection{Approach Overview} \label{sec:approach-ovw}

We illustrate BARO, our proposed end-to-end approach for anomaly detection and root cause analysis for microservice systems based on multivariate time series metrics data, in Fig. \ref{fig:intro2}. BARO consists of two main components. The first component is a Multivariate Bayesian Online Change Point Detection (BOCPD) module to model the dependency and correlation structure of multivariate time-series metrics data so as to detect anomalies (failures). The second component is Robust Scorer, a nonparametric statistical hypothesis testing technique to identify the root cause associated with the failures. Thus, the outputs of BARO include: (1) a boolean indicating whether an anomaly is presented, (2) a ranked list of root cause metrics and their corresponding services, with the highest-ranked items having the highest probability of being the root cause of the failure, if an anomaly is detected. The pseudocode of BARO is described in Algorithm \ref{alg:baro}.

\subsection{Multivariate Bayesian Online Change Point Detection} \label{sec:multi-bocpd}

Anomaly detection in microservices involves identifying anomalies, i.e., observable symptoms of failures \cite{Soldani2022rcasurvey}, while failures in microservices can be considered as interventions that change the monitoring data distribution \cite{Azam2022rcd, Li2022Circa}. Therefore, to detect anomalies (failures) within microservices via time series metrics data, we formulate this problem as a change point detection problem whose goal is to identify whether the behavior of a time series changes significantly. We then propose to use Multivariate BOCPD, a combination of BOCPD \cite{adams2007bayesian} and MultivariateCPD \cite{xuan2007modeling}, to model the dependency and correlation among metrics to detect anomalies effectively. The motivation behind this design is twofold. First, we propose to use BOCPD \cite{adams2007bayesian} as a base technique for detecting change points as it is a simple yet effective online detection technique and it requires no user-specified thresholds to identify change points for univariate time series. It has been shown to be among the best current change point detection methods in many real-world scenarios \cite{Burg2022CPeval}. Second, by combining BOCPD with MultivariateCPD \cite{xuan2007modeling}, we can model the structure and dependency among the multivariate time series metrics data better. This is especially useful for detecting anomalies within microservices due to the failure propagation chain in microservices described in Section \ref{sec:assumption-failure-propagation}. Specifically, anomalies in microservices are generally propagated across the metrics data, causing correlated and dependent changes among different time series metrics. MultivariateCPD has been shown to be able to effectively detect change points when the changes occur in the correlation structure as in multivariate time series metrics data. In the following paragraphs, we describe in detail these two components of our proposed method.

The main idea of BOCPD is to model the \textit{run length}, i.e. the number of consecutive data points in the same distribution, since the last change point, given the data observed so far. Specifically, the run length $r_t$ at time $t$ is defined as $0$ if there is a change point at time $t$, and as $r_{t-1}+1$ otherwise. Given the time series metrics data $\mathcal{M}^{i,j=1:n,1:m}_{t_0:t}$, using the Bayes theorem, the posterior probability distribution of the run length $p(r_t \vert \mathcal{M}^{i=1:n,j=1:m}_{t_0:t})$ can be computed as \cite{adams2007bayesian},
\begin{equation} \label{eq:bocpd-run-length}
    p(r_t \vert \mathcal{M}^{i,j=1:n,1:m}_{t_0:t}) = \frac{\sum\nolimits_{r_{t-1}} p(r_t \vert r_{t-1}) p(\mathcal{M}^{i,j=1:n,1:m}_{t} \vert r_{t-1}, (\mathcal{M}^{i,j=1:n,1:m}_{t})^{(r)}) p(r_{t-1} \vert \mathcal{M}^{i,j=1:n,1:m}_{t_0:t-1})} {p(\mathcal{M}^{i,j=1:n,1:m}_{t_0:t})},
\end{equation}
where $(\mathcal{M}^{i,j=1:n,1:m}_{t})^{(r)}$ denotes the set of observed data points associated with the run $r_t$. The formula in Eq. (\ref{eq:bocpd-run-length}) is recursive, meaning that we can compute the posterior distribution of the run length $r_t$ based on the posterior distribution of $r_{t-1}$, the conditional prior of run length $p(r_t \vert r_{t-1})$ and the distribution of the metrics data $p(\mathcal{M}^{i,j=1:n,1:m}_{t_0:t})$. As suggested in \cite{adams2007bayesian}, the marginal likelihood of the metrics data $p(\mathcal{M}^{i,j=1:n,1:m}_{t_0:t})$ can be chosen as a distribution from the exponential family and the conditional prior of run length $p(r_t \vert r_{t-1})$ can be set based on a hazard function with discrete exponential (geometric) distribution. At each time step $t$, the most probable run length is computed as the value with the highest probability $p(r_t \vert \mathcal{M}^{i,j=1:n,1:m}_{t_0:t})$. Finally, the change points are identified as the data points at the time steps whose run lengths decrease.

The main idea of MultivariateCPD, given the time series metrics data $\mathcal{M}^{i,j=1:n,1:m}_{t_0:t_0+T}$, is to model the metrics data at each data point $\mathcal{M}^{i,j=1:n,1:m}_t$ using a multivariate model \cite{xuan2007modeling}. A common choice is to use the multivariate Gaussian, i.e., $\mathcal{M}^{i,j=1:n,1:m}_t \sim \mathcal{N}(0, \Sigma)$, with $\Sigma$ is an inverse Wishart prior $\Sigma \sim IW(N_0, V_0)$ and $N_0$ is set to be $mn$ which is the number time series within the metrics dataset and $V_0$ is set to be $\hat{\sigma}^2 I$, $I$ is the identity matrix and $\hat{\sigma}$ is the mean of the empirical variance pooled across all the metrics data. With this formulation, let us denote $h=(t_2-t_1)+1$, the marginal likelihood of the multivariate time series data $\mathcal{M}^{i,j=1:n,1:m}_{t_1:t_2}$ can then be computed explicitly as~\cite{xuan2007modeling},
\begin{equation}
\begin{aligned}
p(\{\mathcal{M}^{i,j=1:n,1:m}_{t_1:t_2}\}) &= \pi^{-\frac{hmn}{2}} \frac{\vert V_0 \vert^{N_0/2}}{\vert V_h\vert^{(N_0+h)/2}}  \frac{\Gamma_{mn}(N_0/2)^{-1}}{\Gamma_{mn}((N_0+h)/2)^{-1}}, \\
V_h &= V_0 + S, S=\sum_{i=t_1}^{t_2}  \mathcal{M}^{i,j=1:n,1:m}_t {\big(\mathcal{M}^{i,j=1:n,1:m}_t\big)}^\intercal,
\end{aligned}
\end{equation}
where $\Gamma_{mn}(.)$ denotes the multivariate gamma function. This formulation of the marginal likelihood $p(\mathcal{M}^{i,j=1:n,1:m}_{t_1:t_2})$ can then incorporated into Eq. (\ref{eq:bocpd-run-length}) to replace the univariate marginal likelihood. Other steps are kept the same in order to detect change points in the time series metrics data.

\begin{figure}[t]
\includegraphics[width=0.97\textwidth]{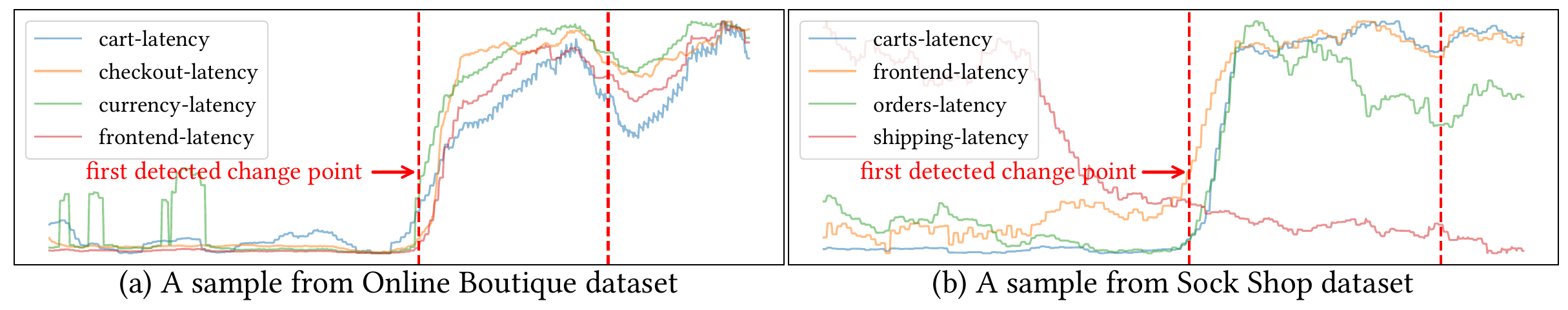}
\vspace{-0.2cm}
\caption{An example of using Multivariate BOCPD to detect change points on multivariate time series data. \textcolor{red}{Dotted vertical red lines} indicate change points. We observe that Multivariate BOCPD can provide the anomaly detection time (the first change point) accurately that separate the normal and abnormal period. In the abnormal period there are multiple change points due to the failure propagation chain.} 
\label{fig:mbocpd}
\vspace{-0.1cm}
\end{figure}

Fig. \ref{fig:mbocpd} presents two examples of using Multivariate BOCPD to detect change points within multivariate time series metrics data using two different datasets. It can be seen that Multivariate BOCPD can accurately detect change points (data points that separate the normal and abnormal data), and thus, detect the failures.

Finally, note that following the assumptions in Section \ref{sec:assumption-causal}, we use \textit{Latency} and \textit{Errors} to detect anomalies, and we only output the first detected change point as the detected anomaly.

\begin{algorithm}[hbt!]
\caption{Pseudo-code of BARO} \label{alg:baro}
\begin{algorithmic}[1]
\Require a set of metrics data $\mathcal{M} = \{x^{i=1:n,j=1:m}_{t_0:t_0+T}\}$ 
\Ensure ranked candidate root causes $\mathcal{R}$, ($\mathcal{R} \in \mathcal{M}$) if detected anomaly

\Function{MultivariateBOCPD}{$\mathcal{M}$}
  \State $\mathcal{M}' \gets$ 
 select \textit{Latency} and \textit{Errors} from $\mathcal{M}$
  \State compute the run length probability $p(r_t \vert \{\mathcal{M'}_{t_0:t}\})$, $\forall x^{(i,j)}\in \mathcal{M'}$, $\forall t\in [t_0, t_0 + T]$
  \State $s_t \gets$ argmax $p(r_t \vert \{\mathcal{M'}_{t_0:t}\}$, $\forall t\in [t_0, t_0 + T]$
  \For{$t\in [t_0, t_0 + T]$} 
    \If {$s_t \le s_{t - 1}$} 
      \State \Return $y = 1, \hat{t}_A = t$; \quad // Returns anomaly if there is a change point detected 
    \EndIf  
  \EndFor   
  \State \Return $y=0,\hat{t}_A=nul$; \quad // Returns normal if there is no change point detected
\EndFunction

\Function{RobustScorer}{$\mathcal{M}$, $\hat{t}_A$}
  \State $\mathcal{R} \gets$ empty list
  \For{$x^{(i, j)} \in \mathcal{M}$}
    \State $\rho^{(i,j)} \gets 0$; med, IQR $\gets$ learn from $\{x^{(i,j)}_{t'}\}$, where $t_0 \le t' \le \hat{t}_A$ 
    \For{$t''$ from $\hat{t}_A$ to $t_{0} + T$}
        \State $a^{(i,j)}_{t''}$ = abs($x^{(i,j)}_{t''}$ - med) / IQR; $\rho^{(i,j)} = \max(\rho^{(i,j)}, a^{(i,j)}_{t''})$
    \EndFor 
    \State $\mathcal{R} \gets \mathcal{R}$ appends ($x^{(i,j)}$, $\rho^{(i,j)}$)
  \EndFor 
  \State $\mathcal{R} \gets$ reversely sort $\mathcal{R}$ based on $\rho^{(i,j)}$
  \State \Return $\mathcal{R}$
\EndFunction

\Procedure{BARO}{$\mathcal{M}$}
  \State $y, \hat{t}_A \gets$ \Call{MultivariateBOCPD}{$\mathcal{M}$} \quad // Step 1: Run Multivariate BOCPD
  
  \State \Return \Call{RobustScorer}{$\mathcal{M}$, $\hat{t}_A$} \textbf{if} $y=1$ \textbf{else} \Return nul // Step 2: Run RobustScorer with $\hat{t}_A$
  
\EndProcedure 
\end{algorithmic}
\end{algorithm}

\subsection{RobustScorer: A Robust Nonparametric Hypothesis Testing Technique} \label{sec:robustscorer}

To identify the root cause metrics, we aim to identify the metrics that exhibit significant changes in data distribution at the anomaly detection time \cite{Shan2019Ediagnosis, liu2023pyrca, Li2022Circa}. To solve this problem, one approach is to conduct hypothesis testing and test whether the data distribution of the metrics data changes significantly after the anomaly detection time. This approach was employed in $\epsilon$-Diagnosis \cite{Shan2019Ediagnosis, liu2023pyrca} and NSigma \cite{Li2022Circa} and they have been shown to perform very well in various scenarios. Our key insight is that previous works might be extremely sensitive to the anomaly detection output (failure occurrence time $\hat{t}_A$), i.e., inaccurate specification of the failure occurrence time might yield bad root cause analysis. Therefore, we propose RobustScorer to address this problem.

Specifically, for each metric $x_{t_0:t_0+T}^{(i,j)}$ in the metrics dataset $\mathcal{M}^{i,j=1:n,1:m}_{t_0:t_0+T}= \{x^{i=1:n,j=1:m}_{t_0:t_0+T}\}$, we conduct a hypothesis test based on the following null hypothesis ($\textbf{H}_0$):\textit{ $x^{(i,j)}$ is not a root cause metric for the failure}. This null hypothesis means that, for $t$ after the anomaly detection time, $x^{(i,j)}_t \sim \mathcal{L}(x^{(i,j)}_{t_\text{normal}})$ with $\mathcal{L}(x^{(i,j)}_{t_\text{normal}})$ denoting the distribution of the metrics data during the normal period (when $t$ is before the anomaly detection time).

This approach generally requires the specification of the anomaly detection time. Inaccurate anomaly detection time therefore could impact the accuracy of these techniques significantly. In this section, we propose a novel nonparametric hypothesis testing technique that is robust and less sensitive to the accuracy of the anomaly detection time.

\subsubsection{A Robust Nonparametric Hypothesis Test}

RobustScorer follows a statistical approach to learn the expected distribution from the metrics data. For every time series $x^{(i,j)}_{t_0:t_0+T}$ in the metrics dataset $\mathcal{M}^{i,j=1:n,1:m}_{t_0:t_0+T}$, let us denote $\hat{t}_A$ as the anomaly detection time, which is an estimation of the time when the anomaly occurs, RobustScorer is trained using the data collected prior to the anomaly, spanning from $t_0$ to $\hat{t}_A$, to learn the median ($med$) and interquartile range ($IRQ$) of this data distribution. Subsequently, for each data point $x^{(i,j)}_t$ in the anomalous period (from $\hat{t}_A$ to $t_0+T$), RobustScorer measures how significant it deviates from the expected central tendency. This deviation is denoted as $a^{(i,j)}_t$ and is computed as follows,
\begin{equation} \label{eq:robust-scoring-1}
    a^{(i,j)}_t = \big | x^{(i,j)}_t - med \big | / IQR.
\end{equation}

All the values of $a^{(i,j)}_t$ across all the metrics data during the anomalous period are then consolidated to yield $\rho^{(i, j)}$, which is an indicator measuring the changes of each metric during the anomalous period,

\begin{equation} \label{eq:robust-scoring-2}
    \rho^{(i, j)} = \max_{\hat{t}_A \leq t \leq t_0+T} a^{(i, j)}_t.
\end{equation}

A higher $\rho^{(i, j)}$ signifies a greater likelihood that $x^{(i, j)}$ serves as the root cause metric, thereby identifying $s^i$ as the root cause service. Finally, RobustScorer then generates a ranked list of root cause metrics based on the magnitudes of $\rho^{(i, j)}$, 
arranged in descending order, with the highest ones corresponding to the most probable root cause metrics (fine-grained root causes). The coarse-grained ranked list of root cause services can be derived from the fine-grained ranking list by extracting the services corresponding to the metrics. Note that we do not use the p-value to reject/accept a possible root cause as in standard hypothesis testing. In our method, we use hypothesis testing to rank the potential root causes as it is normal for the system operators to focus on the top candidates.

Similar to \cite{Li2022Circa, Shan2019Ediagnosis}, our proposed RobustScorer is also distribution-free, scale-equivalent (i.e. metrics data in different scales do not affect the ranked list of root causes), and rotation-invariant (i.e. timestamp shifts do not affect the ranked list of root causes).

\begin{figure}[ht]
\centering
\includegraphics[width=0.98\textwidth]{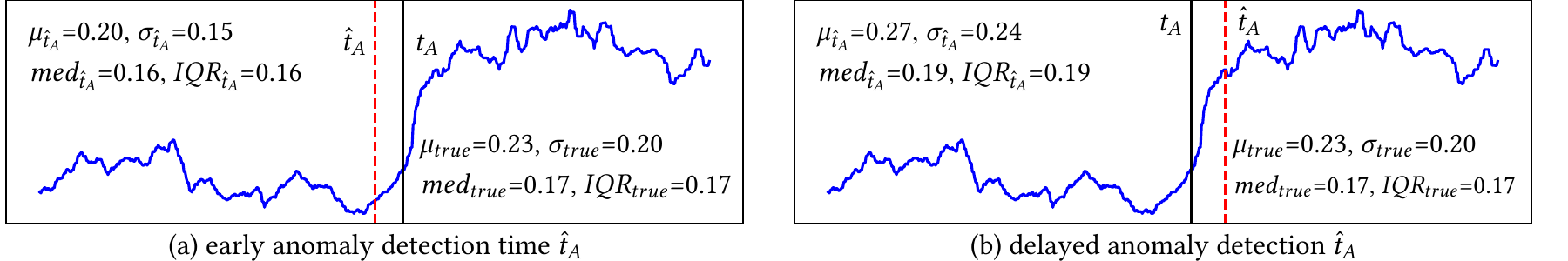}
\vspace{-0.2cm}
\caption{The Robustness of RobustScorer against imprecise anomaly detection time. In (a), an early anomaly detection time reduces the number of data points used to compute the distribution of the normal data in the hypothesis test. Median and IQR show greater resilience to a limited data setting compared to mean and standard deviation. In (b), a delayed anomaly detection time includes abnormal data (outliers) into the normal period. Median and IQR also show robustness to these outliers better than mean and standard deviation.} 
\label{fig:robustsccorer-robustness}
\vspace{-0.2cm}
\end{figure}

\subsubsection{Why is RobustScorer Robust to Imprecise Anomaly Detection Time?} \label{sec:why-robust}

In contrast to previous works \cite{Li2022Circa, Shan2019Ediagnosis, Ma2019Msrank}, which compare normal and abnormal metrics data using the mean and standard deviation of the data distribution, we propose to use the median and interquartile range in our hypothesis testing module. The rationale is that mean and standard deviation are known to be sensitive to outliers, which could be introduced by inaccurate anomaly detection. On the other hand, the median and interquartile range are notably resilient to the impact of outliers \cite{modifiedzscore, robustzscore}.

In Fig. \ref{fig:robustsccorer-robustness}, we illustrate the robustness to the anomaly detection time of RobustScorer corresponding to two scenarios: early anomaly detection (Fig. \ref{fig:robustsccorer-robustness}a) and delayed anomaly detection (Fig. \ref{fig:robustsccorer-robustness}b). An early anomaly detection reduces the number of data points used to compute the distribution of normal data in the hypothesis test. In the scenario of limited data, median and IQR are known to be more resilient than mean and standard deviation, making RobustScorer work well in this scenario. On the other hand, a delayed anomaly detection includes abnormal data (outliers) into the normal data period. Median and IQR are known to work well in the presence of outliers, making RobustScorer to also work well in this scenario. For instance, with RobustScorer, the computed median and IQR values of normal data distribution based on the anomaly detection time $\hat{t}_A$ remain close to those computed based on the true anomaly occurrence time $t_A$. In contrast, the computed mean and standard deviation based on the anomaly detection time $\hat{t}_A$ are more different compared to those computed based on $t_A$. In our experimental evaluation in Section \ref{sec:results}, we demonstrate that RobustScorer outperforms existing RCA approaches in identifying the failure's root cause and generally is more robust to the anomaly detection time than other baselines.

\subsubsection{Handling of Correlated Failures}
For correlated failures affecting multiple services simultaneously, RobustScorer ranks affected services/metrics as the top possible root causes. This allows quicker identification of actual root causes, instead of troubleshooting all possible root causes. For example, consider a firewall misconfiguration leading to correlated failures affecting services A and B simultaneously. Although the true root cause is the misconfiguration, RobustScorer ranks services A and B as the top probable root cause services since they are the immediate successors of the true root cause. The operator can check these services and analyze the true root cause promptly.

\section{Evaluation}

This section answers the following research questions:
\begin{itemize}
    \item \textbf{\textit{RQ1: How effective is BARO in anomaly detection?}} To answer this RQ, we conduct an experiment to compare BARO with state-of-the-art anomaly detection approaches and evaluate their performance in detecting anomalies.
    \item \textbf{\textit{RQ2: How effective is BARO in root cause analysis?}} To answer this RQ, we compare BARO with state-of-the-art RCA methods and evaluate their performance in ranking both root cause services (coarse-grained) and root cause metrics (fine-grained) of the failures.
    \item \textit{\textbf{RQ3: How effective are the components of BARO?}} To answer this RQ, we evaluate the effectiveness of each main component of BARO: Multivariate BOCPD in detecting anomalies and RobustScorer in locating the failure's root cause.
    \item \textbf{\textit{RQ4: How sensitive different RCA methods are w.r.t different parameters?}} To answer this RQ, we perform a sensitivity analysis to evaluate the peformance of all these methods with different values of the anomaly detection time and other methods hyperparameters.
\end{itemize}
    
\label{sec:results}

\subsection{Benchmark Microservice Systems \& Data Collection}

We deploy three well-known benchmark microservice systems, namely Online Boutique \cite{ob}, Sock Shop \cite{sockshop}, and Train Ticket \cite{tt}, on a Kubernetes cluster consisting of one master node and five worker nodes. Each node has 16 CPUs and 32GB RAM, resulting in a total of 80 CPUs and 160GB RAM across five workers. We sequentially deployed the three systems with their default replicas configuration, i.e., one instance per service, to inject faults and gather metrics data under the load of 100-200 concurrent users. Online Boutique is an e-commerce application with 12 services, allowing users to view items, add them to their cart and make purchases. Each service in the Online Boutique system requires between 0.2-0.5 CPU and 64-512MB RAM for normal functioning. Sock Shop is another e-commerce system focused on selling socks, comprising 11 services that communicate via HTTP requests. Each service in the Sock Shop system requires between 0.1-1 CPU and 300MB-2GB RAM for normal functioning. On the other hand, Train Ticket is one of the largest microservice benchmark systems emulating a train ticket booking platform featuring 64 services. Compared to Sock Shop and Online Boutique, Train Ticket has longer and more complex failure propagation paths. Each service in the Train Ticket system requires between 0.2-1 CPU and 200MB-1GB RAM for normal functioning. These benchmark microservice systems have been widely recognized and employed for evaluating the performance of RCA methods \cite{Jinjin2018Microscope, Azam2022rcd, wu2021microdiag, Xin2023CausalRCA, wu2022automatic, he2022graph, dan2021practical, yu2021microrank, zhou2018trainticket, Wang2021evalcausal}.

\begin{figure}[H]
\includegraphics[width=0.8\textwidth]{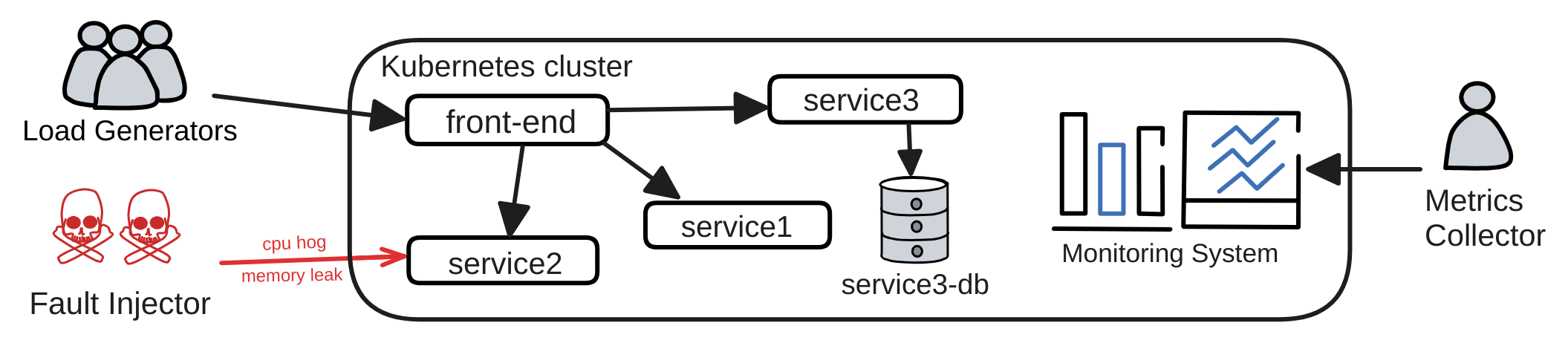}
\vspace{-0.3cm}
\caption{Overview of our setup for microservice systems.} \label{fig:system_setup}
\vspace{-0.3cm}
\end{figure}

To gather the metrics data, we employ the Istio service mesh \cite{istio} along with Prometheus \cite{prometheus} and cAdvisor \cite{cadvisor} to monitor and collect resource-level and service-level metrics, as previously done in~\cite{Azam2022rcd, Xin2023CausalRCA, wu2021microdiag}. To generate traffic, we use the load generators supplied by these systems and tailor them to explore all services with the load of 40-50 requests per second. Fig. \ref{fig:system_setup} illustrates our setup to collect the experimental data from the microservice systems.

We inject four common anomalies: CPU hog, memory leak, network delay, and packet loss into several key services within each benchmark microservice system. Initially, we operate the applications normally %for 10 minutes 
to gather metrics data under normal conditions. Then, we follow the existing practice \cite{Azam2022rcd, lee2023eadro, wu2021microdiag, Xin2023CausalRCA, yu2021microrank} to inject faults into the running services. We execute into the designated container using kubectl exec. For CPU hog and memory leak, we use stress-ng \cite{stressng} to stress the container resource. For network delay and packet loss, we use tc \cite{tc} to manipulate the traffic of the container. Specifically, we inject faults into five targeted services of Sock Shop (carts, catalogue, orders, payment, and user), five targeted services of Online Boutique (adservice, cartservice, checkoutservice, currencyservice, and productcatalogue), and five targeted services of Train Ticket (ts-auth-service, ts-order-service, ts-route-service, ts-train-service, ts-travel-service). We choose these services due to their critical nature, as issues with their performance can impact other services~\cite{Azam2022rcd, Xin2023CausalRCA, lee2023eadro}. For each combination of fault type and targeted service, we repeat the operation (i.e. fault injection and metrics data collection) five times, resulting in 100 failure cases for each benchmark microservice system. 

Furthermore, to ensure the independence of different injection experiments, we chose to restart the microservice systems after each experiment of injecting failures and collecting data, instead of waiting for a cooldown period, as described in previous studies \cite{wu2021microdiag, yu2021microrank, Xin2023CausalRCA}. Upon visual inspections, we observed that the deviations between the fault injection time and the onset of failure symptoms in the metrics data are usually around 1-3 seconds. 
It is important to note that we simulate a workload of 100-200 concurrent users to interact with the systems intensively. Henceforth, the system is highly sensitive to the injected faults, and anomalies are reflected in the metrics shortly after the injection operation. Table \ref{tab:real-data} presents the statistics summarizing the collected data. To the best of our knowledge, we are the first to use all three popular benchmark microservice systems to evaluate the RCA methods. 

\begin{table}
\centering
\caption{Characteristics of collected data from three benchmark microservice systems (\#metrics, \#svc, \#t\_svc: number of metrics, services, and targeted services in the system. \#fault: number of fault types).} \label{tab:real-data}
\vspace{-0.2cm}
\resizebox{0.6\textwidth}{!}{%
\begin{tabular}{l r r r r r r}
\hline
\textbf{Name} & \textbf{\#metrics}  & \textbf{\#svc}   & \textbf{\#t\_svc}  & \textbf{\#fault} & \textbf{\#cases} \\ \hline \hline
Online Boutique & 49 & 12 & 5  & 4 & 100  \\ \hline
Sock Shop & 46 & 11 & 5 & 4 & 100   \\ \hline
Train Ticket & 212 & 64 & 5 & 4 & 100  \\ \hline
\end{tabular}%
}

\end{table}

\subsection{Evaluation Metrics}

\subsubsection{Anomaly Detection}
Similar to previous works \cite{Chen2014Causeinfer, chen2016causeinfer, chen2022adaptive}, we evaluate the anomaly detectors as binary classification models as the main goal is to detect whether there exist anomalies in the metrics data. We determine metrics data collected during a fault injection period as abnormal, while metrics data before that period is considered normal. Therefore, we use Precision, Recall, and F1 scores to evaluate the anomaly detectors. When an anomaly detection algorithm successfully detects an abnormal sample (i.e., a case with anomalies), it is counted as a True Positive (TP). Conversely, incorrectly classifying an abnormal sample as normal is considered False Negative (FN). Likewise, incorrectly classifying a normal sample as abnormal is considered False Positive (FP). The formulas for computing the metrics Precision, Recall, and F1-score are as follows:
\begin{equation}
    Precision = \frac{TP}{TP + FP}, \quad Recall = \frac{TP}{TP + FN}, \quad F1 = \frac{2\times Precision \times Recall}{Precision + Recall}.
\end{equation}

\subsubsection{Root Cause Analysis}

Following existing works \cite{Wang2018cloudranger, Ma2019Msrank, Ma2020Automap, Meng2020Microcause, yu2021microrank, wu2020microrca, wu2021microdiag, Azam2022rcd, Li2022Circa, Xin2023CausalRCA}, we use two standard metrics, namely $AC@k$ and $Avg@k$ to assess the performance of the RCA methods. Herein, we set $k = 1, 3, 5$. Given a set of failure cases A, $AC@k$ and $Avg@k$ are calculated as follows,
\begin{equation}
    AC@k = \frac{1}{|A|} \sum\nolimits_{a\in A}\frac{\sum_{i<k}R^a[i]\in V^a_{rc}}{min(k, |V^a_{rc}|)}, \quad Avg@k = \frac{1}{k}\sum_{j=1}^k AC@j.
\end{equation}
where $R^a[i]$ denotes the $i$th ranking result for the failure case $a$ by an RCA method, and $V^a_{rc}$ is the true root cause set of case $a$. $AC@k$ represents the probability the top $k$ results given by a method include the real root causes. It ranges from $0$ to $1$, with higher values indicating better performance.
Meanwhile, $Avg@k$ measures the overall performance of RCA methods.

% Following existing works \cite{Wang2018cloudranger, Ma2019Msrank, Ma2020Automap, Meng2020Microcause, yu2021microrank, wu2020microrca, wu2021microdiag, Azam2022rcd, Li2022Circa, Xin2023CausalRCA}, we use two standard metrics, namely $AC@k$ and $Avg@k$ to assess the performance of the RCA methods. Herein, we set $k = 1, 3, 5$. Given a set of failure cases A, $AC@k$ is calculated as follows,
% \begin{equation}
%     AC@k = \frac{1}{|A|} \sum\nolimits_{a\in A}\frac{\sum_{i<k}R^a[i]\in V^a_{rc}}{min(k, |V^a_{rc}|)},
% \end{equation}
% where $R^a[i]$ denotes the $i$th ranking result for the failure case $a$ by an RCA method, and $V^a_{rc}$ is the true root cause set of case $a$. $AC@k$ represents the probability the top $k$ results given by a method include the real root causes. It ranges from $0$ to $1$, with higher values indicating better performance.
% $Avg@k$, which measures the overall performance of RCA methods, is calculated as follows,
% \begin{equation}
%     Avg@k = \frac{1}{k}\sum_{j=1}^k AC@j.
% \end{equation}

\subsection{Experimental Setting}
We conduct all the experiments on Linux servers equipped with 8 CPU and 16GB RAM. To avoid any randomness, when evaluating each method, for each failure case we repeat the experiment (detecting anomalies and identifying root causes) five times, then report the average results. Our framework is implemented using Python 3.10.

\vspace{-0.2cm}
\subsection{Baselines} \label{sec:baselines}

\paragraph{Anomaly Detection Baselines.} We select the following four baselines to compare against our proposed method: \textit{N-Sigma} \cite{Jinjin2018Microscope, Li2022Circa}, \textit{BIRCH} \cite{zhang1996birch, wu2020microrca, wu2021microdiag}, \textit{SPOT} \cite{siffer2017anomaly, pan2021dycauserca, Li2022Circa, lee2023eadro, Meng2020Microcause}, and \textit{Univariate Bayesian Change Point Detection} (denoted as UniBCP) \cite{adams2007bayesian, Chen2014Causeinfer, chen2016causeinfer}. These methods are commonly used in existing RCA research for anomaly detection. Their source code has been made publicly available, with the exception of UniBCP, for which we leverage an available implementation \cite{unibcp} for the evaluation. The detailed descriptions of these methods are in Section \ref{sec:background-ad}. For all methods, we use the hyperparameter settings as in previous works and in their public source code.

\vspace{-0.2cm}
\paragraph{Root Cause Analysis Baselines.} We choose six representative baselines, including five state-of-the-art metric-based RCA methods: CausalRCA \cite{Xin2023CausalRCA}, RCD \cite{Azam2022rcd}, CIRCA \cite{Li2022Circa}, $\epsilon$-Diagnosis \cite{Shan2019Ediagnosis}, and N-Sigma \cite{Li2022Circa}, for performance comparison with our proposed method, BARO. Detailed information of these methods is as follows:

\begin{itemize}
\item \textit{Dummy:} Dummy randomly selects a metric as the root cause. We use this method to assess whether our BARO framework and the other baselines outperform random selection.

\item \textit{CausalRCA \cite{Xin2023CausalRCA}:} CausalRCA uses DAG-GNN \cite{Yu2019DagGNN}, a gradient-based causal structure learning method, to estimate the causal graph and uses PageRank algorithm to rank the root causes.

\item \textit{$\epsilon$-Diagnosis \cite{Shan2019Ediagnosis}:}
 $\epsilon$-Diagnosis uses the two-sample test algorithm and $\epsilon$-statistics to estimate the similarity between every pair of metrics and rank the root causes based on test scores.

\item \textit{RCD \cite{Azam2022rcd}:} RCD employs a divide-and-conquer strategy to divide the input metrics into smaller chunks. It uses the $\Psi$-PC algorithm \cite{Jaber2020PsiFCI} to construct a causal graph and identify root causes within each chunk. Subsequently, it combines these root causes and repeats this process until only one chunk remains.

\item \textit{CIRCA \cite{Li2022Circa}:} CIRCA creates a causal graph by using a provided call graph, which requires operational knowledge. Then, it performs regression-based hypothesis testing to find the root causes. Since the call graph is not available, thus following \cite{Azam2022rcd, liu2023pyrca}, we use the PC algorithm to construct this graph for CIRCA. 

\item \textit{N-Sigma \cite{Jinjin2018Microscope,Li2022Circa}:} N-Sigma is a statistical analysis technique that compares the metrics data before and after the anomaly detection time using z-score. The higher the score, the more likely that metric is the root cause of the failure. 

\end{itemize}

The source code of CausalRCA, RCD, CIRCA, and N-Sigma is publicly available. For CIRCA, since the required call graphs are unavailable, we use the PC algorithm to construct the causal graphs as done in \cite{Azam2022rcd, Ma2019Msrank, Ma2020Automap, Wang2018cloudranger, Xin2023CausalRCA}. For $\epsilon$-Diagnosis, the original source code is unavailable thus we rely on the implementation of a related work \cite{liu2023pyrca}. We use the same hyperparameter values as reported in their papers.

\subsection{RQ1: Effectiveness in Anomaly Detection}

In this RQ, we evaluate the performance of BARO and the baseline anomaly detectors across all three datasets. We use the metrics data collected before the failure as normal data, and during the failure as abnormal data. We report the average of Precision (Pre), Recall (Rec), and F1-score (F1) over all the cases. Table \ref{tab:ad1} presents the experimental results, with the best results highlighted in \textbf{bold}. We draw the following observations:

\textbf{(1) BARO consistently outperforms all baseline methods in detecting anomalies by a large margin across all three benchmark microservice systems.} It achieves the highest Precision, Recall, and F1-score on all the datasets. BARO's performance can be attributed to the following factors: (1) BOCPD's ability to detect distribution changes, suitable for identifying anomalies from the failure, acting as a soft intervention \cite{Azam2022rcd, Li2022Circa}, (2) Multivariate BOCPD considers not only individual time series but also the dependencies among the time series, allowing it to detect correlation changes, and (3) it does not require any user-defined thresholds making it to be highly adaptive to different types of metrics data.

\begin{table}
\centering
\caption{Precision, Recall, and F1-score of five anomaly detectors on three datasets: Online Boutique, Sock Shop, and Train Ticket. The best scores are in \textbf{bold}. Higher values indicate better performance.}
\vspace{-0.2cm}
\label{tab:ad1}
\resizebox{0.68\textwidth}{!}{%
\begin{tabular}{l|rrr|rrr|rrr}
\hline
 & \multicolumn{3}{c|}{\textbf{Online Boutique}} & \multicolumn{3}{c|}{\textbf{Sock Shop}} & \multicolumn{3}{c}{\textbf{Train Ticket}} \\ \cline{2-10} 
Method & \multicolumn{1}{c}{\textit{Pre}} & \multicolumn{1}{c}{\textit{Rec}} & \multicolumn{1}{c|}{\textit{F1}} & \multicolumn{1}{c}{\textit{Pre}} & \multicolumn{1}{c}{\textit{Rec}} & \multicolumn{1}{c|}{\textit{F1}} & \multicolumn{1}{l}{\textit{Pre}} & \multicolumn{1}{l}{\textit{Rec}} & \multicolumn{1}{c}{\textit{F1}}  \\ \hline
N-Sigma & 0.54 & \textbf{1} & 0.7 & 0.56 & \textbf{1} & 0.72 & 0.5 & 1 & 0.67 \\ \hline
SPOT & 0.53 & \textbf{1} & 0.69 & 0.53 & \textbf{1} & 0.69 & 0.5 & 1 & 0.67 \\ \hline
BIRCH & 0.35 & 0.48 & 0.4 & 0.05 & 0.05 & 0.05 & 0.39 & 0.4 & 0.39 \\ \hline
UniBCP & 0.51 & \textbf{1} & 0.67 & 0.5 & 0.99 & 0.66 & 0.5 & 1 & 0.67 \\ \hline
\textbf{BARO (Ours)} & \textbf{0.69} & \textbf{1} & \textbf{0.82} & \textbf{0.6} & \textbf{1} & \textbf{0.75} & \textbf{0.68} & \textbf{1} & \textbf{0.81} \\ \hline
\end{tabular}%
}

{\footnotesize \textit{(*) UniBCP denotes Univariate Offline Bayesian Change Point Detection.}}

\vspace{-0.2cm}
\end{table}

\textbf{(2) All baseline methods, except BIRCH, have very high recalls but low precisions across all the datasets.} This indicates that they generally can detect anomalies, however, they also frequently misclassify normal time series as anomalous. The precision of these methods generally range from 0.05 to 0.68. Based on our observations, this is due to the complexity and dynamics of the benchmark microservice systems, which include many time series metrics with complicated patterns, leading to frequent misclassifications of anomalies by these baseline methods.

In summary, BARO stands out as the best-performing method, being effective in detecting anomalies. This anomaly detection capability plays a crucial role in the subsequent RCA task.

\subsection{RQ2: Effectiveness in Root Cause Analysis}

In this section, we evaluate the performance of BARO and the RCA baseline methods on all three datasets. First, we assess the ideal performance of the methods when the anomaly occurrence time is known accurately. Specifically, we use the fault injection time $t_{\text{inject}}$ as a proxy for the true anomaly occurence time since we have observed that in our benchmark microservice systems, the anomaly occurence time closely aligns with the fault injection time. Then, we assess the performance of RCA methods when using the anomaly detection time provided by existing anomaly detectors described in Section \ref{sec:baselines}. Note that Dummy and CausalRCA do not require the specification of the anomaly detection time. Tables \ref{tab:q2-rca-s} and \ref{tab:q2-rca-f1} report the overall performance of all methods using the Avg@5 scores for coarse-grained root cause analysis (root cause services) and fine-grained root cause analysis (root cause metrics), respectively. We calculate the accuracy for each type of fault: CPU hog (CPU), memory leak (MEM), network delay (DELAY), and packet loss (LOSS), as well as their average (AVG) to report the overall performance across four fault types. In the tables, we denote the combination of an RCA method and an anomaly detector as \textit{RCA\_method[Anomaly\_Detector]}. For instance, RCD[N-Sigma], RCD[BIRCH], RCD[SPOT], RCD[UniBCP], and RCD[BOCPD] represent the RCA pipelines that combine RCD with the anomaly detectors N-Sigma, BIRCH, SPOT, Univariate BCPD, and Multivariate BOCPD, respectively.

\begin{table}[]
\centering
\caption{Coarse-grained performance of different RCA methods in terms of Avg@5 on three different datasets. The fault types CPU, MEM, DELAY, and LOSS denote CPU overload, memory leak, network delay, and packet loss. The highest scores are in \textbf{bold}, and the second highest scores are in {\ul underscore}.}
\label{tab:q2-rca-s}
\vspace{-0.2cm}
\resizebox{\textwidth}{!}{%
\setlength\tabcolsep{3pt}
\begin{tabular}{l|rrrrr|rrrrr|rrrrr}
\hline
 & \multicolumn{5}{c|}{\textbf{Online Boutique}} & \multicolumn{5}{c|}{\textbf{Sock Shop}} & \multicolumn{5}{c}{\textbf{Train Ticket}} \\ \cline{2-16} 
Method & \multicolumn{1}{c}{\textit{CPU}} & \multicolumn{1}{c}{\textit{MEM}} & \multicolumn{1}{c}{\textit{DELAY}} & \multicolumn{1}{c}{\textit{LOSS}} & \multicolumn{1}{c|}{\textit{AVG}} & \multicolumn{1}{c}{\textit{CPU}} & \multicolumn{1}{c}{\textit{MEM}} & \multicolumn{1}{c}{\textit{DELAY}} & \multicolumn{1}{c}{\textit{LOSS}} & \multicolumn{1}{c|}{\textit{AVG}} & \multicolumn{1}{c}{\textit{CPU}} & \multicolumn{1}{c}{\textit{MEM}} & \multicolumn{1}{c}{\textit{DELAY}} & \multicolumn{1}{c}{\textit{LOSS}} & \multicolumn{1}{c}{\textit{AVG}} \\ \hline \hline
Dummy & 0.24 & 0.26 & 0.27 & 0.24 & 0.25 & 0.37 & 0.41 & 0.38 & 0.37 & 0.38 & 0.07 & 0.08 & 0.07 & 0.07 & 0.07 \\
CausalRCA & 0.85 & 0.91 & 0.86 & 0.58 & 0.8 & 0.49 & 0.82 & 0.61 & 0.47 & 0.6 & 0.53 & 0.3 & 0.17 & 0.11 & 0.28 \\ \hline
$\epsilon$-Diagnosis {[}$t_{\text{inject}}${]} & 0.1 & 0.13 & 0.12 & 0.12 & 0.12 & 0.47 & 0.3 & 0.42 & 0.49 & 0.42 & 0 & 0.02 & 0 & 0 & 0.01 \\
$\epsilon$-Diagnosis {[}N-Sigma{]} & 0.19 & 0.16 & 0.23 & 0.22 & 0.2 & 0.49 & 0.39 & 0.46 & 0.5 & 0.46 & 0 & 0 & 0 & 0.02 & 0.01 \\
$\epsilon$-Diagnosis {[}BIRCH{]} & 0.23 & 0.33 & 0.21 & 0.14 & 0.22 & - & 0 & - & 1 & 0.8 & 0.07 & 0 & 0 & 0 & 0.02 \\
$\epsilon$-Diagnosis {[}SPOT{]} & 0.37 & 0.2 & 0.17 & 0.22 & 0.24 & 0.46 & 0.34 & 0.42 & 0.46 & 0.42 & 0 & 0.02 & 0 & 0 & 0.01 \\
$\epsilon$-Diagnosis {[}UniBCP{]} & 0.28 & 0.09 & 0.4 & 0.2 & 0.24 & 0.2 & 0.19 & 0.25 & 0.2 & 0.21 & 0.05 & 0.10 & 0.08 & 0.04 & 0.07  \\
$\epsilon$-Diagnosis {[}BOCPD{]} & 0.23 & 0.09 & 0.18 & 0.15 & 0.16 & 0.45 & 0.35 & 0.38 & 0.44 & 0.41 & 0 & 0 & 0 & 0.06 & 0.02 \\ \hline
RCD {[}$t_{\text{inject}}${]} & 0.69 & 0.4 & 0.27 & 0.5 & 0.47 & 0.62 & 0.46 & 0.48 & 0.37 & 0.48 & 0.08 & 0.01 & 0.09 & 0.12 & 0.08 \\
RCD {[}N-Sigma{]} & 0.67 & 0.43 & 0.31 & 0.49 & 0.48 & 0.54 & 0.42 & 0.46 & 0.37 & 0.45 & 0.07 & 0 & 0.05 & 0.1 & 0.06 \\
RCD {[}BIRCH{]} & 0.69 & 0.37 & 0.28 & 0.50 & 0.46 & 0.57 & 0.42 & 0.49 & 0.37 & 0.46 & 0.06 & 0 & 0.03 & 0.11 & 0.05 \\
RCD {[}SPOT{]} & 0.7 & 0.42 & 0.3 & 0.48 & 0.48 & 0.58 & 0.43 & 0.46 & 0.34 & 0.45 & 0.08 & 0.01 & 0.07 & 0.09 & 0.06 \\
RCD {[}UniBCP{]} & 0.66 & 0.42 & 0.29 & 0.49 & 0.47 & 0.62 & 0.44 & 0.5 & 0.36 & 0.48 & 0.05 & 0.01 & 0.05 & 0.1 & 0.05 \\
RCD {[}BOCPD{]} & 0.7 & 0.4 & 0.3 & 0.52 & 0.48 & 0.6 & 0.47 & 0.48 & 0.36 & 0.48 & 0.06 & 0 & 0.04 & 0.09 & 0.05 \\ \hline
CIRCA {[}$t_{\text{inject}}${]} & {\ul 0.9} & 0.74 & 0.9 & 0.55 & 0.77 & {\ul 0.97} & \textbf{0.98} & {\ul 0.98} & {\ul 0.88} & {\ul 0.95} & 0.66 & 0.93 & 0.64 & 0.57 & 0.7 \\
CIRCA {[}N-Sigma{]} & 0.78 & 0.59 & 0.79 & 0.48 & 0.66 & 0.86 & 0.7 & 0.87 & 0.7 & 0.78 & 0.65 & 0.88 & 0.58 & 0.57 & 0.67 \\
CIRCA {[}BIRCH{]} & 0.45 & 0.38 & 0.27 & 0.48 & 0.39 & - & 0.60 & - & 0.75 & 0.72 & 0.10 & 0 & 0.35 & 0.13 & 0.19 \\
CIRCA {[}SPOT{]} & 0.76 & 0.7 & 0.88 & 0.66 & 0.75 & 0.83 & 0.87 & 0.95 & 0.78 & 0.86 & 0.69 & 0.96 & 0.58 & 0.53 & 0.69 \\
CIRCA {[}UniBCP{]} & 0.16 & 0 & 0 & 0.35 & 0.13 & 0.23 & 0 & 0.2 & 0.36 & 0.2 & 0 & 0 & 0 & 0.33 & 0.08 \\
CIRCA {[}BOCPD{]} & 0.68 & 0.2 & 0.28 & 0.42 & 0.4 & 0.73 & \textbf{0.98} & 0.54 & 0.66 & 0.73 & 0.37 & 0.12 & 0.3 & 0.17 & 0.24 \\ \hline
N-Sigma {[}$t_{\text{inject}}${]} & {\ul 0.9} & 0.93 & {\ul 0.94} & 0.66 & \textbf{0.86} & \textbf{0.98} & \textbf{0.98} & {\ul 0.98} & \textbf{0.9} & \textbf{0.96} & 0.81 & 0.96 & 0.61 & \textbf{0.7} & 0.77 \\
N-Sigma {[}N-Sigma{]} & 0.79 & 0.69 & 0.83 & 0.63 & 0.74 & 0.85 & 0.77 & 0.89 & 0.79 & 0.83 & 0.74 & \textbf{0.98} & 0.61 & 0.65 & 0.75 \\
N-Sigma {[}BIRCH{]} & 0.83 & 0.78 & 0.64 & 0.6 & 0.71 & 		
- & 0 & - & 0.85 & 0.68 & 0.55 & 0.7 & 0.47 & 0.45 & 0.51 \\
N-Sigma {[}SPOT{]} & 0.79 & 0.79 & 0.93 & 0.65 & 0.79 & 0.87 & 0.91 & \textbf{0.99} & 0.82 & 0.9 & 0.75 & \textbf{0.98} & 0.61 & 0.64 & 0.75 \\
N-Sigma {[}UniBCP{]} & 0.75 & 0.69 & 0.91 & {\ul 0.76} & 0.78 & 0.86 & 0.79 & 0.69 & 0.81 & 0.79 & 0.38 & 0.71 & 0.38 & 0.62 & 0.52 \\
N-Sigma {[}BOCPD{]} & 0.74 & 0.32 & 0.52 & 0.58 & 0.54 & 0.8 & \textbf{0.98} & 0.63 & 0.62 & 0.76 & 0.51 & 0.16 & 0.29 & 0.22 & 0.3 \\ \hline 
RobustScorer {[}$t_{\text{inject}}${]} & {\ul 0.9} & {\ul 0.94} & {\ul 0.94} & 0.65 & \textbf{0.86} & \textbf{0.98} & \textbf{0.98} & {\ul 0.98} & 0.86 & {\ul 0.95} & \textbf{0.9} & {\ul 0.97} & {\ul 0.62} & {\ul 0.67} & {\ul 0.79} \\
RobustScorer {[}N-Sigma{]} & {\ul 0.9} & \textbf{0.96} & 0.88 & 0.58 & 0.83 & 0.96 & \textbf{0.98} & {\ul 0.98} & \textbf{0.9} & \textbf{0.96} & 0.82 & \textbf{0.98} & {\ul 0.62} & 0.64 & 0.77 \\
RobustScorer {[}BIRCH{]} & 0.92 & 0.93 & 0.81 & 0.51 & 0.79 & - & 1 & - & 0.85 & 0.88 & 0.76 & 0.55 & 0.57 & 0.71 & 0.65 \\
RobustScorer {[}SPOT{]} & 0.89 & {\ul 0.94} & 0.93 & 0.61 & {\ul 0.84} & {\ul 0.97} & \textbf{0.98} & \textbf{0.99} & 0.86 & {\ul 0.95} & {\ul 0.85} & \textbf{0.98} & 0.61 & 0.66 & 0.78 \\
RobustScorer {[}UniBCP{]} & 0.73 & 0.67 & 0.89 & \textbf{0.79} & 0.77 & 0.8 & 0.65 & 0.69 & 0.84 & 0.75 & 0.37 & 0.64 & 0.38 & 0.54 & 0.48 \\ \hline
\textbf{BARO (Ours)} & \textbf{0.91} & \textbf{0.96} & \textbf{0.95} & 0.62 & \textbf{0.86} & {\ul 0.97} & \textbf{0.98} & {\ul 0.98} & 0.87 & {\ul 0.95} & \textbf{0.9} & \textbf{0.98} & \textbf{0.64} & \textbf{0.7} & \textbf{0.81} \\ \hline
\end{tabular}%
}

{\footnotesize \textit{(*) BIRCH has a poor recall, so RCA results with BIRCH are obtained from a limited number of failure cases where BIRCH detects anomalies.
CIRCA results for TrainTicket are only obtained from 15/100 cases due to PC's failed causal graph construction.
}}
\vspace{-0.3cm}
\end{table}

% (*) BIRCH has a poor recall, so the results of RCA methods combined with BIRCH are calculated based on a limited number of failure cases.
% CIRCA results for Train Ticket are only obtained from 15/100 cases due to PC's failed causal graph construction.

\subsubsection{Coarse-grained Root Cause Analysis}

From Table \ref{tab:q2-rca-s}, we draw the following observations.

\textbf{(1) BARO performs the best in identifying the coarse-grained root cause of the failure}, consistently achieving top accuracy scores for all types of faults across all datasets. BARO achieves the highest Avg@5 in 10 out of 15 cases and outperforms other baselines by a large margin. For example, while CausalRCA achieves 0.8, 0.6, and 0.28, BARO achieves 0.86, 0.95, and 0.81 in the overall Avg@5 for all three microservice systems, respectively. This represents 7.5\%, 58\%, and 189\% improvements compared to CausalRCA.

From the results, we can see that, when working with large-scale microservice systems, $\epsilon$-Diagnosis, RCD, CausalRCA, and CIRCA encounter huge difficulties, whereas BARO yields much better results. BARO's resistance to the imprecision of the anomaly detection time via the nonparametric RobustScorer hypothesis test demonstrates its applicability in practical scenarios, where the anomaly detection module may differ across real-world systems. Moreover, BARO leverages multivariate BOCPD to capture dependencies within multivariate metrics data, allowing for more precise localization of the anomaly occurrence time and thus improve the RCA performance.

\textbf{(2) CausalRCA does not require separating the normal and abnormal metrics data.} It uses all the provided metrics data to construct the causal graph and then uses PageRank to rank the root causes. The experimental results indicate that it performs well on the Online Boutique system. For instance, it achieves an Avg@5 of 0.85 for the CPU overload fault, 0.91 for the memory leak fault, and an overall Avg@5 of 0.8. Nevertheless, its performance declines on the other two systems. For example, it only achieves overall Avg@5 scores of 0.6 and 0.28 on the Sock Shop and Train Ticket systems, possibly due to the challenges posed by the complex structures of these systems. Overall, we can see that the performance of CausalRCA varies significantly across different systems.

\textbf{(3) $\epsilon$-Diagnosis, a widely-known baseline RCA method, does not perform greatly superior to random selection.} For example, its overall Avg@5 scores range from 0.12 to 0.24, whereas random selection achieves 0.25 on the Online Boutique system. However, it outperforms Dummy by 20\% in the Sock Shop system, with an overall Avg@5 score of 0.46 compared to Dummy's 0.38.

\textbf{(4) RCD outperforms random selection in small-scale systems such as Online Boutique and Sock Shop.} For example, its overall Avg@5 scores range from 0.45 to 0.48 while Dummy's scores range from 0.25 to 0.38. However, for the large-scale Train Ticket system, RCD's performance drops remarkably. For instance, RCD's Avg@5 scores range from 0.05 to 0.08, similar to Dummy's. This observation suggests the need to include large-scale systems like Train Ticket in future work to benchmark newly proposed RCA methods.

\textbf{(5) We observe that N-Sigma, a simple statistical analysis technique, performs better than CIRCA.} Note that, in this work, as discussed, for CIRCA, we use the causal graphs constructed by the PC algorithm instead of the required call graphs as these graphs are not available for these systems, which could be a cause of this observation. Overall, N-Sigma outperforms both CIRCA and RCD in coarse-grained root cause localization.

In summary, BARO consistently outperforms other methods for all fault types and datasets by a large margin. When applied to a large-scale system like Train Ticket, baseline methods like $\epsilon$-Diagnosis, RCD, CausalRCA, and CIRCA struggle, while BARO still delivers superior results.

\begin{table}[!t]
\centering
\caption{Fine-grained performance of different RCA methods in terms of Avg@5 accuracy on three different datasets. The fault types CPU, MEM, DELAY, and LOSS denote CPU overload, memory leak, network delay, and packet loss. The highest scores are in \textbf{bold}, and the second highest scores are in {\ul underscore}.}
\label{tab:q2-rca-f1}
\vspace{-0.2cm}
\resizebox{\textwidth}{!}{%
\setlength\tabcolsep{3pt}
\begin{tabular}{l|rrrrr|rrrrr|rrrrr}
\hline
 & \multicolumn{5}{c|}{\textbf{Online Boutique}} & \multicolumn{5}{c|}{\textbf{Sock Shop}} & \multicolumn{5}{c}{\textbf{Train Ticket}} \\ \cline{2-16} 
Method & \multicolumn{1}{c}{\textit{CPU}} & \multicolumn{1}{c}{\textit{MEM}} & \multicolumn{1}{c}{\textit{DELAY}} & \multicolumn{1}{c}{\textit{LOSS}} & \multicolumn{1}{c|}{\textit{AVG}} & \multicolumn{1}{c}{\textit{CPU}} & \multicolumn{1}{c}{\textit{MEM}} & \multicolumn{1}{c}{\textit{DELAY}} & \multicolumn{1}{c}{\textit{LOSS}} & \multicolumn{1}{c|}{\textit{AVG}} & \multicolumn{1}{c}{\textit{CPU}} & \multicolumn{1}{c}{\textit{MEM}} & \multicolumn{1}{c}{\textit{DELAY}} & \multicolumn{1}{c}{\textit{LOSS}} & \multicolumn{1}{c}{\textit{AVG}} \\ \hline \hline
Dummy & 0.08 & 0.06 & 0.07 & 0.06 & 0.07 & 0.1 & 0.07 & 0.07 & 0.06 & 0.08 & 0.02 & 0.02 & 0.02 & 0.01 & 0.02 \\
CausalRCA & \textbf{0.55} & \textbf{0.78} & 0.86 & 0.39 & \textbf{0.65} & 0.39 & 0.42 & 0.49 & 0.34 & 0.41 & 0.51 & 0.13 & 0.1 & 0.09 & 0.21 \\ \hline
$\epsilon$-Diagnosis {[}$t_{\text{inject}}${]} & 0.07 & 0.03 & 0.06 & 0.02 & 0.05 & 0 & 0 & 0 & 0 & 0 & 0 & 0 & 0 & 0 & 0 \\
$\epsilon$-Diagnosis {[}N-Sigma{]} & 0 & 0.06 & 0.2 & 0.13 & 0.1 & 0.02 & 0 & 0 & 0 & 0.01 & 0 & 0 & 0 & 0 & 0 \\
$\epsilon$-Diagnosis {[}BIRCH{]} & 0 & 0.07 & 0.21 & 0.14 & 0.11 &  - & 0 & - & 0 & 0 & 0 & 0 & 0.2 & 0 & 0 \\
$\epsilon$-Diagnosis {[}SPOT{]} & 0.06 & 0.02 & 0.14 & 0.06 & 0.07 & 0 & 0 & 0 & 0 & 0 & 0 & 0 & 0 & 0 & 0 \\
$\epsilon$-Diagnosis {[}UniBCP{]} & 0 & 0 & 0.3 & 0.12 & 0.11 & 0 & 0 & 0.11 & 0.09 & 0.05 & 0.05 & 0 & 0.05 & 0 & 0.03 \\
$\epsilon$-Diagnosis {[}BOCPD{]} & 0.02 & 0.04 & 0.1 & 0.12 & 0.07 & 0 & 0 & 0 & 0 & 0 & 0 & 0 & 0 & 0 & 0 \\ \hline
RCD {[}$t_{\text{inject}}${]} & 0.16 & 0.26 & 0.22 & 0.18 & 0.21 & 0.02 & 0.17 & 0.1 & 0.07 & 0.09 & 0 & 0 & 0.06 & 0 & 0.02 \\
RCD {[}N-Sigma{]} & 0.16 & 0.29 & 0.23 & 0.15 & 0.21 & 0 & 0.16 & 0.09 & 0.07 & 0.08 & 0 & 0 & 0.03 & 0 & 0.01 \\
RCD {[}BIRCH{]} & 0.15 & 0.27 & 0.24 & 0.19 & 0.21 & 0.01 & 0.15 & 0.11 & 0.06 & 0.08 & 0 & 0 & 0.02 & 0 & 0 \\
RCD {[}SPOT{]} & 0.14 & 0.31 & 0.23 & 0.15 & 0.21 & 0.02 & 0.16 & 0.11 & 0.06 & 0.09 & 0 & 0 & 0.04 & 0 & 0.01 \\
RCD {[}UniBCP{]} & 0.16 & 0.31 & 0.22 & 0.14 & 0.21 & 0.01 & 0.16 & 0.09 & 0.07 & 0.08 & 0 & 0 & 0.04 & 0 & 0.01 \\
RCD {[}BOCPD{]} & 0.17 & 0.28 & 0.24 & 0.17 & 0.22 & 0.04 & 0.16 & 0.14 & 0.07 & 0.1 & 0 & 0 & 0.03 & 0 & 0.01 \\ \hline
CIRCA {[}$t_{\text{inject}}${]} & 0.42 & 0.3 & 0.9 & {\ul 0.54} & 0.54 & 0.52 & 0.58 & {\ul 0.98} & {\ul 0.86} & 0.74 & 0.48 & 0.86 & 0.55 & 0.57 & 0.62 \\
CIRCA {[}N-Sigma{]} & 0.22 & 0.27 & 0.79 & 0.4 & 0.42 & 0.5 & 0.42 & 0.87 & 0.66 & 0.61 & 0.48 & 0.84 & 0.51 & 0.54 & 0.59 \\
CIRCA {[}BIRCH{]} & 0.17 & 0 & 0.29 & 0.38 & 0.23 & - & 0 & - & 0.75 & 0.60 & 0.10 & 0 & 0.31 & 0.13 & 0.18 \\
CIRCA {[}SPOT{]} & 0.34 & 0.18 & 0.88 & 0.6 & 0.5 & 0.5 & 0.46 & 0.94 & 0.74 & 0.66 & 0.5 & 0.88 & 0.5 & 0.5 & 0.6 \\
CIRCA {[}UniBCP{]} & 0 & 0 & 0 & 0.2 & 0.05 & 0.08 & 0 & 0 & 0.15 & 0.06 & 0 & 0 & 0 & 0.17 & 0.04 \\
CIRCA {[}BOCPD{]} & 0.29 & 0 & 0.18 & 0.35 & 0.21 & 0.42 & 0.49 & 0.34 & 0.47 & 0.43 & 0.3 & 0.12 & 0.18 & 0.09 & 0.17 \\ \hline
N-Sigma {[}$t_{\text{inject}}${]} & {\ul 0.54} & 0.54 & {\ul 0.94} & 0.5 & {\ul 0.63} & 0.77 & 0.7 & {\ul 0.98} & \textbf{0.88} & {\ul 0.83} & \textbf{0.6} & 0.93 & 0.54 & \textbf{0.65} & \textbf{0.68} \\
N-Sigma {[}N-Sigma{]} & 0.38 & 0.33 & 0.83 & 0.48 & 0.51 & 0.66 & 0.46 & 0.87 & 0.66 & 0.66 & 0.55 & {\ul 0.94} & 0.55 & 0.59 & 0.66 \\
N-Sigma {[}BIRCH{]} & 0.38 & 0.33 & 0.63 & 0.46 & 0.45 & - & 0 & - & 0.85 & 0.68 & 0.33 & 0.65 & 0.39 & 0.33 & 0.38 \\
N-Sigma {[}SPOT{]} & 0.41 & 0.41 & 0.92 & 0.53 & 0.57 & 0.66 & 0.58 & \textbf{0.99} & 0.77 & 0.75 & {\ul 0.56} & {\ul 0.94} & 0.55 & 0.58 & 0.66 \\
N-Sigma {[}UniBCP{]} & 0.37 & 0.11 & 0.91 & 0.66 & 0.51 & 0.51 & 0.08 & 0.59 & 0.74 & 0.48 & 0.2 & 0.39 & 0.38 & {\ul 0.62} & 0.4 \\
N-Sigma {[}BOCPD{]} & 0.46 & 0.04 & 0.51 & 0.46 & 0.37 & 0.6 & 0.54 & 0.42 & 0.52 & 0.52 & 0.35 & 0.16 & 0.21 & 0.14 & 0.22 \\ \hline 
RobustScorer {[}$t_{\text{inject}}${]} & 0.51 & 0.48 & {\ul 0.94} & 0.49 & 0.61 & \textbf{0.81} & 0.68 & {\ul 0.98} & 0.79 & 0.82 & 0.54 & 0.9 & {\ul 0.57} & 0.6 & 0.65 \\
RobustScorer {[}N-Sigma{]} & {\ul 0.54} & {\ul 0.56} & 0.88 & 0.49 & 0.62 & {\ul 0.8} & {\ul 0.72} & {\ul 0.98} & 0.83 & {\ul 0.83} & 0.46 & \textbf{0.96} & 0.52 & 0.57 & 0.63 \\
RobustScorer {[}BIRCH{]} & 0.42 & 0.51 & 0.79 & 0.43 & 0.54 & - & 0.8 & - & 0.85 & 0.84 & 0.33 & 0.45 & 0.49 & 0.67 & 0.49 \\
RobustScorer {[}SPOT{]} & {\ul 0.54} & 0.51 & 0.93 & 0.47 & 0.61 & \textbf{0.81} & \textbf{0.78} & \textbf{0.99} & 0.78 & \textbf{0.84} & 0.5 & 0.93 & 0.52 & 0.58 & 0.63 \\
RobustScorer {[}UniBCP{]} & 0.27 & 0.04 & 0.89 & \textbf{0.61} & 0.45 & 0.46 & 0.08 & 0.59 & 0.73 & 0.47 & 0.13 & 0.31 & 0.38 & 0.54 & 0.34 \\ \hline
\textbf{BARO (Ours)} & 0.51 & 0.51 & \textbf{0.95} & 0.47 & 0.61 & 0.79 & 0.67 & {\ul 0.98} & 0.8 & 0.81 & 0.54 & 0.93 & \textbf{0.58} & 0.61 & {\ul 0.67} \\ \hline
\end{tabular}%
}
\vspace{-0.2cm}
\end{table}

\subsubsection{Fine-grained Root Cause Analysis} From Table \ref{tab:q2-rca-f1}, we can see that \textbf{BARO consistently ranks among the top performers}, with overall Avg@5 scores of 0.61, 0.81, and 0.67, while the top scores among all the methods are 0.65, 0.84, and 0.68 on Online Boutique, Sock Shop, and Train Ticket systems, respectively. Notably, BARO significantly outperforms RCD and CIRCA. When provided $t_{\text{inject}}$, RCD achieves overall Avg@5 of 0.21, 0.09, and 0.02, CIRCA reaches 0.54, 0.74, and 0.62, while BARO achieves 0.61, 0.81, and 0.67 on Online Boutique, Sock Shop, and Train Ticket, respectively. BARO improves baseline RCA methods' performance from 8\% to 800\%. While BARO performs similarly to CausalRCA on Online Boutique, it outperforms CausalRCA significantly on the Sock Shop and Train Ticket systems, with improvements ranging from 97\% to 219\% in overall Avg@5 score, demonstrating its capability in working with complex systems. These consistent results establish BARO as a reliable method across different microservice systems, demonstrating BARO's superior performance compared to recent RCA state-of-the-art approaches \cite{Xin2023CausalRCA, Azam2022rcd, Li2022Circa}. 

\subsection{RQ3: Effectiveness of BARO's Components.} \label{exp:q3-robustness}

In this section, we analyze the performance of major components of BARO to assess their contribution to the overall pipeline. The experimental results in Tables \ref{tab:q2-rca-s} and \ref{tab:q2-rca-f1} also support this section. We derive the following observations based on the experimental results:

(1) \textbf{Multivariate BOCPD presents superior performance among the anomaly detectors when integrated with RobustScorer.} While different anomaly detectors can be combined with RobustScorer to provide good performance, Multivariate BOCPD stands out by consistently delivering the best results. Multivariate BOCPD enables BARO to achieve the top overall Avg@5 score (0.86 on Online Boutique, 0.95 on Sock Shop, and 0.81 in Train Ticket for coarse-grained RCA). Multivariate BOCPD is effective because it can detect distribution changes within multivariate time series, making it suitable for detecting anomalies for microservices. N-Sigma, a simple anomaly detector, can enhance the RCA performance but its best performance %when combined with all the RCA methods 
is still lower than BARO. For example, on Online Boutique, its best performance is 0.83 whilst BARO achieves 0.86. BIRCH has poor recalls (i.e. detects only 48/100 abnormal cases on Online Boutique), limiting its ability to support RCA. When it can detect anomalies, its performance is still much lower than other anomaly detection methods. SPOT's performance is similar to N-Sigma, and is lower than BARO. 
Finally, Univariate BOCPD's performance is among the worst when combined with any RCA methods.
%Univariate BOCPD's performance is among the worst anomaly detectors, i.e., when combined with any RCA methods, its performance is always among the worst.

(2) \textbf{RobustScorer demonstrates superior robustness} compared to other baselines when integrated with different anomaly detectors. When provided $t_{\text{inject}}$, all three methods (CIRCA, N-Sigma, and BARO) perform similarly across three benchmark systems. Specifically, on the Online Boutique, Sock Shop, and Train Ticket systems, CIRCA achieves scores of 0.77, 0.95, and 0.7, N-Sigma scores of 0.86, 0.96, and 0.77, and RobustScorer scores of 0.86, 0.95, and 0.79, respectively. However, when we employ different anomaly detection methods (i.e. N-Sigma, BIRCH, SPOT, Univariate BCP, and Multivariate BOCPD) to estimate the anomaly occurrence time, the RCA methods reveal varying degrees of sensitivity. For example, on the Online Boutique system, CIRCA exhibits a variation of 62\%, N-Sigma shows a variation of 33\%, and our RobustScorer displayed a variation of 25\%. 
On Online Boutique, N-Sigma achieves Avg@5 scores of 0.74, 0.46, 0.79, whilst RobustScorer achieves scores of 0.83, 0.61, 0.84, representing improvements of 12\%, 32\%, and 6\%  compared to when using N-Sigma, BIRCH, and SPOT as anomaly detectors, respectively.
Overall, our RobustScorer demonstrated less sensitivity. When combined with Multivariate BOCPD, RobustScorer achieves accuracy similar to those obtained with $t_{\text{inject}}$. This observation, once again, confirms the effectiveness of our method.
 
In summary, the two components of BARO significantly contribute to the approach's effectiveness. They both play critical roles in our proposed root cause analysis pipeline.

\subsection{RQ4: Sensitivity Analysis of RCA Methods} \label{exp:q4-sensitivity}

In this section, we conduct a sensitivity analysis to assess the performance of various RCA methods w.r.t. different critical parameters. We use the AC@1, AC@3, and Avg@5 scores to assess the performance of the methods. We explore two key aspects of this sensitivity analysis.

\subsubsection{Sensitivity on the Anomaly Detection Time $\hat{t}_A$}
\begin{figure}
\vspace{-0.1cm}
\includegraphics[width=0.9\textwidth]{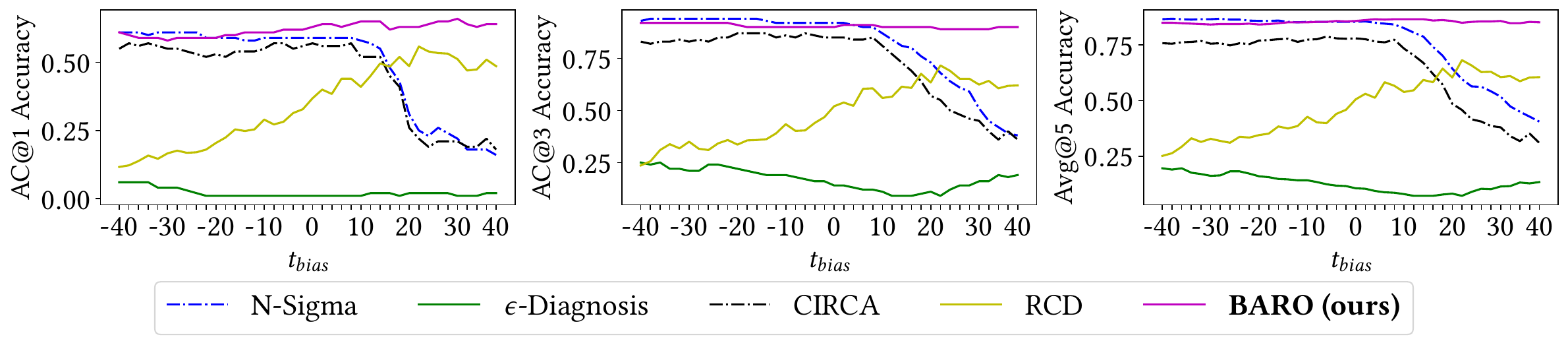}
\vspace{-0.3cm}
\caption{The performance of N-Sigma, $\epsilon$-Diagnosis, CIRCA, RCD, and BARO w.r.t. different values of $t_{\text{bias}}$ on the Online Boutique dataset. The figure presents the AC@1, AC@3, and Avg@5 scores from left to right.} \label{fig:q4-sensitivity}
\vspace{-0.3cm}
\end{figure}

Let us denote $t_{\text{inject}}$ as the fault injection time. Here, we aim to assess the performance of RCA methods when the anomaly detection time varies around this fault injection time. We formulate the anomaly detection time $\hat{t}_{A}$ as $\hat{t}_{A} = t_{\text{inject}} + t_{\text{bias}}$ where $t_{\text{bias}}$ ranges from -40 to 40. We then evaluate the performance of the RCA methods with different anomaly detection time within this range. We run the experiments with the Online Boutique and Sock Shop datasets. The experimental results on the Online Boutique dataset are in Fig. \ref{fig:q4-sensitivity} whilst the results on the Sock Shop dataset are in our supplementary material, Fig. S1 \cite{barogithubzenodo, barogithub}.

We observe that BARO and $\epsilon$-Diagnosis exhibit significant resistance to variations in the specification of $\hat{t}_{A}$. In contrast, the performance of N-Sigma and CIRCA drops significantly when the anomaly is detected late. This issue might stem from the sensitivity of the mean and standard deviation to outliers, as discussed in Section \ref{sec:why-robust}. In particular, both N-Sigma and CIRCA use z-score, $z=\frac{x-\mu}{\sigma}$, which is computed based on the mean and standard deviation of the data distribution before the anomaly detection time. When the anomaly is detected late, a number of anomalous data points is included in the normal data period, and with the use of mean and standard deviation, this leads to improper root cause ranking (see Fig. \ref{fig:robustsccorer-robustness}). Finally,
RCD also has high sensitivity to $\hat{t}_A$, highlighting a weakness of the combination of $\Psi$-PC and the divide-and-conquer strategy. % when dealing with this parameter.

%Finally, RCD also shows a high sensitivity to this value, highlighting the weakness of the combination of $\Psi$-PC and the divide-and-conquer strategy when dealing with this imperfect parameter.

\subsubsection{Hyperparameter Sensitivity}

\begin{figure}
\vspace{-0.2cm}
\includegraphics[width=0.95\textwidth]{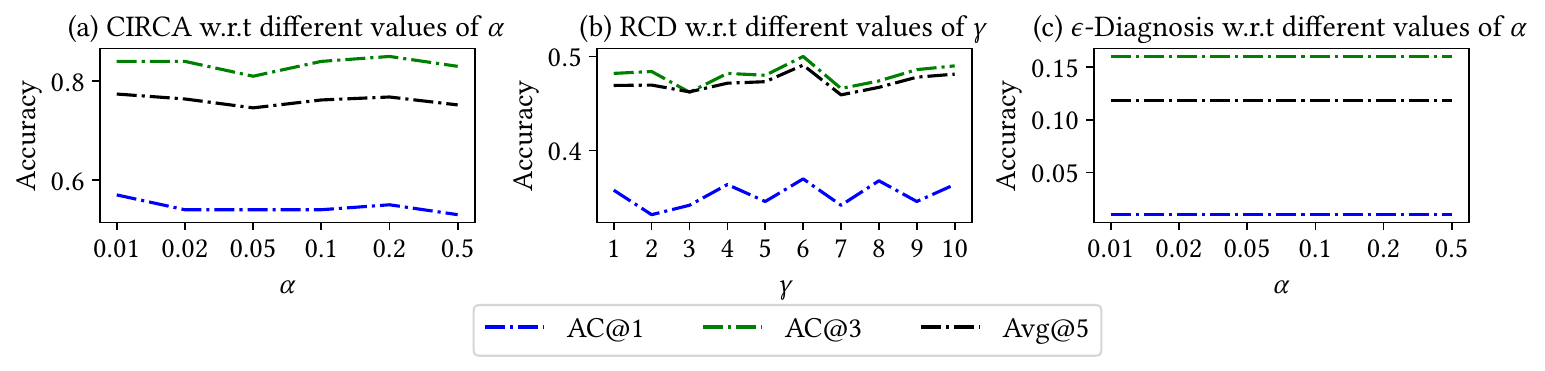}
\vspace{-0.2cm}
\caption{The performance of CIRCA (a), RCD (b), and $\epsilon$-Diagnosis (c) w.r.t. their different parameter values on the Online Boutique dataset. The figure presents their AC@1, AC@3, and Avg@5 scores.} \label{fig:q4-sensitivity-combine}
\vspace{-0.2cm}
\end{figure}

In this section, we aim to assess the sensitivity of different RCA methods to their parameters. RCD employs a divide-and-conquer strategy, requiring the specification of a chunk size parameter denoted as $\gamma$. We vary this parameter from 1 to 10 while the default value is 5. CIRCA uses PC to construct the causal graph, requiring a threshold $\alpha$ for independence testing. We vary this parameter from 0.01 to 0.5 while the default value is 0.05. $\epsilon$-Diagnosis also requires a threshold $\alpha$ to drop the insignificant metrics. We range $\alpha$ from 0.01 to 0.5 in this case.
Note that CausalRCA's parameters are already tuned \cite{Xin2023CausalRCA}, eliminating parameter sensitivities.
Fig. \ref{fig:q4-sensitivity-combine} presents our experimental results on the Online Boutique dataset whilst the experimental results on the Sock Shop dataset are in the supplementary material, Fig. S2 \cite{barogithub, barogithubzenodo}.

% In this section, we aim to assess the sensitivity of different RCA methods to their respective parameters. RCD employs a divide-and-conquer strategy, requiring the specification of a chunk size parameter denoted as $\gamma$. We vary this parameter from 1 to 10 while the default value is 5. CIRCA uses the PC algorithm to construct the causal graph, requiring a threshold $\alpha$ for independence testing. We vary this parameter within the range of 0.01 to 0.5 while the default value is 0.05. $\epsilon$-Diagnosis also requires a threshold $\alpha$ to drop the insignificant metrics. We range $\alpha$ from 0.01 to 0.5 in this case. Note that according to \cite{Xin2023CausalRCA}, CausalRCA's parameters are already tuned, and we employ their best configuration, eliminating the need for parameter sensitivity analysis in this case. Fig. \ref{fig:q4-sensitivity-combine} presents our experimental results on the Online Boutique dataset whilst the experimental results on the Sock Shop dataset are in the supplementary material (Fig. S2) \cite{barogithub, barogithubzenodo}. 

Based on these results, it is evident that RCD, CIRCA, and $\epsilon$-Diagnosis exhibit minimal sensitivity to their parameters. Consequently, we can have confidence in the reliability of the results obtained in our %previous 
research questions. Notably, the performance of $\epsilon$-Diagnosis is identical with different values of $\alpha$. This can be attributed to the fact that in $\epsilon$-Diagnosis, the $\alpha$ parameter primarily influences the trimming of lower-ranked root causes. Given our focus on the top-k results, the length of the list becomes less critical. Similarly, for CIRCA, the main component that serves root cause analysis is its statistical scorer; the significance level $\alpha$ mainly affects the construction of the causal graph, which has minimal impact on the performance. For RCD, slight fluctuations in performance are observed with varying chunk size $\gamma$, but these variations are deemed insignificant.

\subsection{Running Time \& Instrumentation Cost}

% Due to space limit, the running time analysis of different anomaly detection and RCA methods are available in our replication package \cite{barogithub}. Adopting BARO in practice necessitates the installation of monitoring agents alongside application services to gather system and application metrics. Further information on installation and how to use BARO can be found in our replication package \cite{barogithub}.

\subsubsection{Running Time}

On three datasets, BARO takes an average of 30 to 173 seconds to perform anomaly detection and root cause analysis. The highest recorded running time of BARO is 7 minutes in a case of the Train Ticket dataset. RobustScorer is very fast when it takes around 0.01 seconds to analyze the root cause, i.e., at least 3 to 1000 times faster than other methods like $\epsilon$-Diagnosis, RCD, CIRCA, and CausalRCA. The detailed running time of our proposed method and different anomaly detection and root cause analysis modules are shown in Tables \ref{tab:running-time-ad} and \ref{tab:running-time-rca}, respectively.

\begin{table}[ht]
\centering
\begin{minipage}[t]{0.48\textwidth}
\centering
\caption{Average anomaly detection run time (in seconds)}
\vspace{-0.3cm}
\label{tab:running-time-ad}
\resizebox{\textwidth}{!}{%
\begin{tabular}{l|r|r|r}
\hline
Method & \multicolumn{1}{c|}{\textbf{Online Boutique}} & \multicolumn{1}{c|}{\textbf{Sock Shop}} & \multicolumn{1}{c}{\textbf{Train Ticket}} \\ \hline \hline
N-Sigma & 0.16 & 0.11 & 0.53 \\ \hline
BIRCH & 0.05 & 0.04 & 0.11 \\ \hline
SPOT & 3.17 & 1.9 & 11.4 \\ \hline
UniBCD & 819.7 & 638.19 & 3292.48 \\ \hline
\textbf{BARO (Ours)} & 44.83 & 30 & 173.37 \\ \hline
\end{tabular}%
}

% {\footnotesize \textit{(*) Univariate Bayesian Online Change Point Detection.}}
\end{minipage}
\hfill
\begin{minipage}[t]{0.48\textwidth}
\centering
\caption{Average RCA run time (in seconds)}
\vspace{-0.2cm}
\label{tab:running-time-rca}
\resizebox{\textwidth}{!}{%
\begin{tabular}{l|r|r|r}
\hline
Method & \multicolumn{1}{l|}{\textbf{Online Boutique}} & \multicolumn{1}{l|}{\textbf{Sock Shop}} & \multicolumn{1}{l}{\textbf{Train Ticket}} \\ \hline \hline 
CausalRCA & 299.18 & 287.18 & 2638.51 \\ \hline
$\epsilon$-Diagnosis & 3.94 & 3.97 & 14.83 \\ \hline
RCD & 10.74 & 5.62 & 24.21 \\ \hline
CIRCA & 13.52 & 13.47 & 7564.88 \\ \hline
N-Sigma & 0.01 & 0.01 & 0.01 \\ \hline
\textbf{BARO (Ours)} & 0.01 & 0.01 & 0.01 \\ \hline
\end{tabular}%
}
\end{minipage}
\vspace{-0.3cm}
\end{table}

\subsubsection{Instrumentation Cost}

BARO requires monitoring agents (e.g., Istio, cAdvisor, Prometheus) installed alongside application services to collect system-level (e.g., CPU/Mem/Disk usage) and application-level metrics (e.g., response time, request count per minute). These metrics are used for BARO to perform anomaly detection and root cause analysis. BARO requires Python and some standard packages (e.g. pandas, numpy). Further details are available in our artifacts \cite{barogithub, barogithubzenodo}.

\label{sec:discussion}

\section{Threats to validity}

We assess potential threats to the validity of our work, considering the construct, internal, conclusion, and external factors as outlined in \cite{wohlin2012experimentation}. The \textbf{construct threat} primarily concerns hyperparameters settings and evaluation metrics. To address this, we conduct a sensitivity analysis for all methods that require specific parameters and use established evaluation metrics. The \textbf{internal threat} concerns the framework implementation, where bugs may affects the results reliability. To mitigate, we use established Python packages, rigorous testing, and repeat the experiments multiple times. The \textbf{conclusion threat} is tied to the fault types as microservices can experience various faults \cite{mariani2018localizing}. We use four common faults to evaluate and demonstrate BARO's superior performance. Furthermore, our framework relies on a set of assumptions discussed and validated in previous works on microservice systems, as described in Sec \ref{sec:assumption-causal}. If a system meets these assumptions, BARO could be applied. Expanding BARO to work with other types of systems, e.g., distributed database systems, could be a potential future work. The \textbf{external threat} is related to the deployment of microservice applications and data collection strategies. In this paper, we employ a single deployment setting for each microservice system and a single data collection strategy, which potentially limits the generality of our work.  However, we deploy the benchmark systems on the real 5-node Kubernetes clusters and repeat the experiments multiple times to get comprehensive datasets. In addition, we use popular benchmark microservice applications, such as Online Boutique, Sock Shop, and Train Ticket, which are widely recognized in academia for testing microservices-related methods~\cite{Jinjin2018Microscope, Azam2022rcd, wu2021microdiag, Xin2023CausalRCA, wu2022automatic, he2022graph, dan2021practical, yu2021microrank, zhou2018trainticket, Wang2021evalcausal}. Furthermore, we adopt Prometheus, an open-source tool for real-time monitoring, and collect service-level and resource-level metrics that present the status of a running microservice application. This choice helps mitigate this threat.

\section{Conclusion}
\label{sec:conclusion}

This paper proposes BARO, a novel end-to-end approach for anomaly detection and root cause analysis for microservice systems based on multivariate time series metrics data. BARO uses Multivariate Bayesian Online Change Point Detection for anomaly detection and introduces the RobustScorer, a nonparametric statistical hypothesis testing technique, for identifying root causes. Our experimental results on three benchmark systems demonstrate that BARO consistently outperforms the state-of-the-art methods in both anomaly detection and root cause analysis. Comprehensive sensitivity analysis highlights the robustness and broad applicability of our approach. Our research contributes a valuable tool for metric-based root cause analysis of microservice systems. 
In future work, we plan to enhance our framework by incorporating multimodal data (e.g., logs, traces). 
% In future work, we plan to enhance anomaly detection and RCA by incorporating multimodal data (e.g., logs, traces) and integrating new advances in ML/DL (e.g., large language models). 
% We also plan to enhance the RCA's trustworthiness by providing explainable root causes.

\section{Data Availability}
We have open-sourced BARO, which can be accessed on GitHub at \cite{barogithub}. Additionally, an immutable artifact for BARO is available on Zenodo \cite{barogithubzenodo}, together with three experimental datasets \cite{barodatasets}.

% The replication package for this paper, including the source code of BARO, three experimental datasets, and the reproducibility instructions, are publicly available \cite{baroartifacts}. 

\begin{acks}
This research was supported by the Australian Research Council Discovery Project (DP220103044), AWS Cloud Credit for Research, and Google Travel Grant. We also thank anonymous reviewers for their insightful and constructive comments, which significantly improve this paper. 
\end{acks}

\bibliographystyle{ACM-Reference-Format}
\bibliography{reference}

\end{document}